\documentclass[aps,floatfix,prd,showpacs]{revtex4}
\usepackage{mathtools}
\usepackage{amsmath,amssymb,epsfig,bm}
\usepackage[english] {babel}
 \addto{\captionsenglish}{}
\usepackage{amsfonts}
\usepackage{xcolor}
\usepackage{caption}
\usepackage{titlesec}
\usepackage{listings}
\usepackage{placeins}
\usepackage{graphics}
\usepackage{graphicx}
\usepackage{wrapfig}
\usepackage{etoolbox}
\usepackage{xcolor}
\usepackage[normalem]{ulem}
\usepackage{cancel}

\usepackage{placeins}
\usepackage{mathrsfs}
\usepackage{setspace}
\usepackage[utf8]{inputenc}
\usepackage[titletoc,toc,title]{appendix}
\usepackage{wrapfig}
\usepackage{ dsfont }
\usepackage{subfig}
\usepackage{graphicx}
\usepackage{booktabs}
\numberwithin{equation}{section}
\titleformat{\paragraph}
{\normalfont\normalsize\bfseries}{\theparagraph}{1em}{}
\titlespacing*{\paragraph}
{0pt}{3.25ex plus 1ex minus .2ex}{1.5ex plus .2ex}

\begin{document}

\title{Effect of high harmonic loops on gravitational wave bounds from cosmic strings}
\author{Despoina Pazouli$^{1}$, Konstantinos Palapanidis$^{2}$, Anastasios Avgoustidis$^{1}$ and Edmund J. Copeland$^{1}$}
\affiliation{
$^{1}$School of Physics and Astronomy, University of Nottingham, Nottingham NG7 2RD, United Kingdom\\
$^{2}$ Gymnasium of Leonidio, Leonidio 22300, Greece} 
\date{\today}

\begin{abstract}
Based on a binary tree model for the self intersection of cosmic string loops containing high harmonics we estimate the number of  self-intersections of the parent and daughter loops and the associated cusp production to determine the most likely number of cusp events per period on the resultant non-self intersecting loops, and provide an updated calculation for the gravitational wave signal that arrives on Earth from cusps on such loops. 
This is done for different numbers of cusps supported from the cosmic strings of the network, and for different harmonic distributions on the loops. 
We plot our results of the event rate of gravitational waves emanating from the cusps in terms of redshift, having fixed the value of $G\mu$ and the received frequency of the signal, and compare our results to those in \cite{Abbott:2017mem,LIGOScientific:2021nrg}.
\end{abstract}

\maketitle

\pagenumbering{arabic}

\section{Introduction}\label{Intro}
Cosmic strings are line-like topological defects which may have formed in the early universe by symmetry breaking phase transitions and are predicted by a wide range of models~\cite{Kibble:1976sj, Kibble:1980mv,Vilenkin:1984ib,Hindmarsh:1994re,Jeannerot:2003qv,Copeland:2009ga,Copeland:2011dx}. Given that they could have formed in the early universe, depending on the energy scale at which this occurred it would have determined their string tension $\mu$, which provides a natural dimensionless coupling parameter when coupled to gravity $G \mu$. This combination is what is commonly constrained by observations, hence provide insights into the physics of the very early universe, such as the the value of parameters associated with grand unification theories from which strings can emerge \cite{Jeannerot:2003qv}. 

A network of cosmic strings, once formed, consists primarily of long infinite strings stretching across the observable universe, and loops of string. Following an initial period of friction domination, the traditional picture of the network evolution is one where the long string density decreases as they intercommute forming loops which in turn decay primarily due to gravitational wave emission. This is based primarily on what is known as the Nambu-Goto approximation, where the strings are considered as effectively infinitely thin line-like objects evolving under their tension. As a result, the cosmic string network, in the scale of a Hubble volume and at cosmic time t, consists of a number of long (infinite) strings that stretch across the Hubble volume and a significantly larger number of closed loops \cite{vilenkin1994cosmic,anderson2015mathematical}. 

The possible detection of the gravitational waves emitted from the evolving network of loops of string remains a holy grail of early universe cosmology, providing perhaps the first observational signatures of physics of that era, which could also include the first evidence of string theory through observations of cosmic superstrings. Therefore the direct observations of gravitational waves (GW) by the LIGO/VIRGO consortium \cite{LIGOScientific:2016aoc} has renewed interest in searching for the gravitational signal produced by cosmic strings, and has led to two recent LIGO/VIRGO constraint papers  \cite{Abbott:2017mem,LIGOScientific:2021nrg}. The strings emit gravitational waves as they oscillate and this is the main mechanism that leads to their decay, according to the Nambu-Goto model. We will adopt that type of model in this paper, but we need to mention that there is a school of thought that argues the primary decay route for strings is through particle production associated with the fact that cosmic strings are really field theory objects typically modelled as Abelian-Higgs strings, and as such have another mode of decay \cite{Hindmarsh:2017qff,Hindmarsh:2021mnl}. We will describe the gravitational wave signal of cosmic strings produced by cusp events on cosmic string loops. Our approach will be different from other work in one critical feature. Following on from \cite{Pazouli:2020qmj}, we will aim to investigate the impact initial loops containing modes of high harmonics can have on the signal, by introducing a toy model to predict the average number of daughter loops a given parent loop produces, and the number of associated cusps produced on these loops. The result will be the derivation of the signal for classes of cosmic string network models which contain parent loops of high harmonic order in them \cite{Pazouli:2020qmj}. 

The paper is laid out as follows: in Section~\ref{CosmicStringNet} we first discuss the cosmic string networks and define the functions required to describe them, for example as used in the most recent LIGO cosmic strings constraint paper \cite{LIGOScientific:2021nrg}. We then introduce a toy model in section~\ref{toymodelevol} that can be used to provide an analytic approach to describe the splitting of loops through self-intersections with respect to their harmonic order and apply the results to our cosmic string model. In Section~\ref{PropGWBFLRW} we calculate the gravitational wave burst (GWB) signal, i.e. the gravitational wave signal from cusps on cosmic strings, for the parameter values that we derived with our toy model. In section~\ref{GWBcs} we calculate the rate of GWBs from the cosmic string network determined by our model, and then in  section~\ref{results} we compare the results based on our choice of parameters with the LIGO results of \cite{Abbott:2017mem,LIGOScientific:2021nrg}. Finally we conclude in section~\ref{concl}. 
\section{Cosmic string network}\label{CosmicStringNet}
A network of cosmic strings evolving from the friction dominated period consists of a distribution of long (infinite) strings stretching across the observable universe and loops, which were either chopped off the long strings as they self-intercommute or were originally produced as the network formed during the phase transition responsible for it (for details see Refs. \cite{vilenkin1994cosmic}, \cite{Copeland:2009ga} and \cite{Binetruy:2012ze}). As the network evolves, new loops are constantly chopping off the long strings, and decaying primarily via gravitational radiation. Although much is known about the network and its properties, many key aspects remain unknown and have to be estimated based either on empirical arguments or simulations. For example, what is the typical size of a loop formed compared to the the Hubble scale ($\sim t$) at the time of formation, what do typical loops look like in terms of the number of harmonics contained on them, how many cusps are produced as they oscillate and how many kinks does a typical loop contain, what is the effect of gravitational backreaction on the evolution of the network, and how does it affect the string energy density scaling with the background energy density? They are important questions, their solutions lead to key parameters which play crucial roles in determining the observational consequences of cosmic strings. Our goal here is to try and reduce the uncertainty in a few of these parameters, and in doing so, show the impact they can have in estimating the amount of the gravitational radiation emitted from cusp events on loops of cosmic strings. Given the uncertainty in the typical loop size produced, we will follow earlier work and consider the Model 1 that was introduced in the LIGO collaboration paper \cite{Abbott:2017mem}. For a recent review of the loop distribution function see Ref.~\cite{Blanco-Pillado:2019tbi}. A key assumption of this model is that all loops chopped off the long string network at a cosmic time $t_i$ are non self-intersecting with period $T_i=l_i/2$ where $l_i$ is the length of the cosmic string loop formed and is given by 
\begin{equation}
l_i \simeq \alpha t_i \, .
\label{looplength}
\end{equation}
Here, $\alpha <1$ is a dimensionless parameter whose precise value is unknown, and depends on the mechanisms that caused the loop to form. For example in Model 1 of \cite{Abbott:2017mem}, they assume $\alpha \sim 0.1$. 

Our main goal is to calculate the event rate of GWBs from loops of cosmic string in such a network, where we also allow for high-harmonic loops to be present and model their evolution. To achieve this we will follow closely the calculation from \cite{Damour:2000wa}, adopting the cosmic string network from \cite{Abbott:2017mem,LIGOScientific:2021nrg}, whilst adding a model representing the evolution of high-harmonic loops. We begin by introducing the following functions for cosmic time and proper distance, which generally have to be evaluated numerically (see section \ref{dRnumer}), as only their asymptotic form can be obtained analytically. The cosmic time is defined as
\begin{equation}
t(z)=\frac{\varphi_t(z)}{H_0}
\label{cosmictimedef}
\end{equation}
where
\begin{equation}
\varphi_t(z)=\int_z^{\infty}\frac{dz'}{\mathcal{H}(z')(1+z')},
\label{phitdef}
\end{equation}
the proper distance (or cosmic distance) is defined as
\begin{equation}
r(z)=\frac{\varphi_r(z)}{H_0}
\label{cosmicdistdef}
\end{equation}
where
\begin{equation}
\varphi_r(z)=\int_0^{z}\frac{dz'}{\mathcal{H}(z')},
\label{phirdef}
\end{equation}
and the proper spatial volume between redshifts $z$ and $z+dz$ is
\begin{equation}
dV(z)=\frac{\varphi_V(z)}{H_0^3}dz
\label{dVzdef}
\end{equation}
where
\begin{equation}
\varphi_V(z)=\frac{4\pi\varphi_r^2(z)}{(1+z)^3\mathcal{H}(z)}.
\label{phiVdef}
\end{equation}
In the above, we have expressed the Hubble parameter at redshift z as
\begin{equation}
H(z)=H_{0}\mathcal{H}(z)
\label{Hconstdef}
\end{equation}
where, in terms of the density parameters for the cosmological constant ($\Omega_{\Lambda}$), radiation ($\Omega_{R}$) and matter ($\Omega_{M}$)
\begin{equation}
\mathcal{H}(z)=\sqrt{\Omega_{\Lambda}+\Omega_{M}(1+z)^3+\Omega_{R}\mathcal{G}(z)(1+z)^4},
\label{Hcaldef}
\end{equation}
with $\Omega_{\Lambda}=1-\Omega_{M}-\Omega_{R}$. The radiation-matter equality redshift is estimated to be $z_{eq}=3366$. We will use the same values for the cosmological parameters as used in \cite{Abbott:2017mem}. These are the Planck 2015 results presented in \cite{Ade:2015xua}, with $H_0=100h\,km\,s^{-1}\,Mpc^{-1}$, $h=0.678$, $\Omega_{M}=0.308$, $\Omega_R=9.1476\times10^{-5}$
The function $\mathcal{G}(z)$ is related to the entropy released from the particle species when they become non-relativistic as the universe cools. It varies mainly during the epochs of electron-positron annihilation and the QCD phase transition \cite{Binetruy:2012ze}, and can be approximated by the piecewise-function
\begin{equation}
\mathcal{G}(z)= \begin{cases} 
      1 & z< 10^9, \\
      0.83 & 10^9< z<2\times 10^{12},  \\
      0.39 & 2\times 10^{12}< z. 
   \end{cases}
\label{Gcaldef}
\end{equation}

The integrals for $\varphi_t(z)$ \eqref{phitdef} and $\varphi_r(z)$ \eqref{phirdef} cannot be evaluated analytically for general $z$, and need to be computed numerically. However, in the limit $z\gg 1$, deep in the radiation dominated era we can simplify the Hubble parameter \eqref{Hcaldef} yielding 
\begin{equation}
\mathcal{H}(z)\simeq \sqrt{\Omega_{R}}z^2.
\label{Hcalzgg1}
\end{equation}
Then, using \eqref{Hcalzgg1}, the integral \eqref{phitdef} can easily be calculated
\begin{equation}
\varphi_t(z)=\frac{1}{2\sqrt{\Omega_R\mathcal{G}(z)}}\frac{1}{z^2},
\label{phitgg1}
\end{equation}
for $z\gg1$.
We cannot use the above method to calculate the large $z$ limit of $\phi_{r}$, because the limits of integration in (\ref{phirdef}) extend beyond the range of validity of the approximation (\ref{Hcalzgg1}), all the way down to $z=0$. Numerically we find that for $z \gg 1$, we obtain the constant solution
\begin{equation}
\varphi_r(z)\simeq 3.39.
\label{phirgg12}
\end{equation}
For $z\ll 1$, we can estimate $\varphi_t$ to be
\begin{equation}
\varphi_t(z)=\int_0^{\infty}\frac{dz'}{\mathcal{H}(z')(1+z')}.
\label{phitll1}
\end{equation}
Upon calculating the above integral, we find the small $z$ value of $\varphi_t(z)$ is roughly constant $\simeq 0.96$. Finally, to calculate the small $z$ behavior of $\varphi_r(z)$, we perform a Taylor series expansion of the integrand function in \eqref{phirdef} around $z=0$, obtaining to leading order 
\begin{equation}
\varphi_r(z)=z,
\label{phirll1}
\end{equation}
for $z\ll 1$.
 
In what follows we will adopt a large loop scenario where the parameter $\alpha =0.1$ in equation \eqref{looplength}  \cite{Binetruy:2012ze},\cite{Abbott:2017mem}. The rate of length loss remains the same as in the small loop case, i.e. $dl/dt=-\Gamma G\mu$, implying that large loops live longer than small loops, and moreover they can survive for longer than a Hubble time. This means that we will need to use a different treatment for the large loop network as we need to include the fact their length decreases due to gravitational radiation. The parameter $\Gamma \sim 50$ is determined from numerical simulations of the decay of loops - see for example \cite{vilenkin1994cosmic}. We first define the relative size of a loop of length $l$ compared to the age of the universe, $t$, as
\begin{equation}
\gamma=l/t
\label{gammarelsize}
\end{equation}
and the loop distribution function as 
\begin{equation}
\mathcal{F}(\gamma,t)=n(l,t)t^4
\label{calFdef}
\end{equation}
where $n(l,t)$ is the number density of loops of length $l$ at cosmic time $t$. For simplicity, we will also assume that the loops do not self-intersect \footnote{Later, when we introduce a model for the evolution of high-harmonic self-intersecting loops, we will in effect replace them by an equivalent set of non-intersecting loops with the equivalent number of cusps, as determined by the model. This model will affect only the calculation of the number $c$, the average number of cusps per period.}.  Therefore,  taking into account the length decrease due to the gravitational wave emission, a loop formed at cosmic time $t_i$ will have a length at a later time, $t$, given by   
\begin{equation}
l(t)=\alpha t_{i} -\Gamma G\mu(t-t_{i}).
\label{lengthtime}
\end{equation}

The loop distribution of large loops in the radiation era ($z>3366$) is \cite{vilenkin1994cosmic,Abbott:2017mem}
\begin{equation}
\mathcal{F}_{rad}(\gamma)=\frac{C_{rad}}{\left(\gamma+\alpha\right)^{5/2}}\Theta(\alpha-\gamma)\,,
\label{calFraddef}
\end{equation}
where $C_{rad}$ is a constant specific to the radiation era.
The Heaviside step function ensures that $\gamma<\alpha$, which is always true since all the loops are formed with a length $\alpha t$ and decay throughout their evolution. In the matter era the loop distribution function consists of two different contributions; one from loops formed in the radiation era that survive into the matter era, and one from loops formed it the matter era, i.e. at times $t>t_{eq}$. Loops formed in the radiation era will have length at cosmic time $t>t_{eq}$ which is less than the length of a loop formed at $t_{eq}$, i.e. $l(t)<\alpha t_{eq}-\Gamma G\mu(t-t_{eq})$, since any loop with length larger than this would have formed in the matter era. Thus, if we define the function
\begin{equation}
\beta(t)=\alpha t_{eq}-\Gamma G\mu(t-t_{eq})
\label{betalength}
\end{equation}
we can then write for the loop distribution function contribution which consists of the radiation era loops that survive into the matter era
\begin{equation}
\mathcal{F}_{mat}^{(1)}(\gamma)=\frac{C_{rad}}{\left(\gamma+\alpha\right)^{5/2}}\left(\frac{t_{eq}}{t}\right)^{1/2}\Theta(-\gamma+\beta(t))\,,
\label{calFmat1def}
\end{equation}
while for the contribution to the loop distribution function from loops formed in the matter era, we have
\begin{equation}
\mathcal{F}_{mat}^{(2)}(\gamma)=\frac{C_{mat}}{\left(\gamma+\alpha\right)^{2}}\Theta(\alpha-\gamma)\Theta(\gamma-\beta(t))\,,
\label{calFmat2def}
\end{equation}
where $C_{mat}$ is a constant specific to the matter era. The function $\Theta(\alpha-\gamma)$ ensures that all loops considered have lengths smaller than the formation length at time t, and the function $\Theta(\gamma-\beta(t))$ ensures that no loops surviving from the radiation era are taken into account in the matter era loop distribution. Therefore, the total matter era loop distribution function is
\begin{equation}
\mathcal{F}_{mat}(\gamma)=\mathcal{F}_{mat}^{(1)}(\gamma)+\mathcal{F}_{mat}^{(2)}(\gamma)
\label{calFmatdef}
\end{equation} 
The constants $C_{rad}$ and $C_{mat}$ are obtained numerically and are given by \cite{vilenkin1994cosmic}, \cite{Siemens:2006vk}, \cite{Abbott:2017mem},  
\begin{equation}
C_{rad}\simeq 1.6,\qquad C_{mat}\simeq 0.48.
\label{Cradmat}
\end{equation} 

\section{A toy model for the loop evolution}\label{toymodelevol}
The value of the average number of cusps per loop period, denoted by $c$ in the following, is a key quantity in the calculation of the GWB signals emitted from cusps on cosmic strings, but remains an elusive one due to the many unknown parameters of cosmic string loop initial conditions and evolution. In \cite{Pazouli:2020qmj} we determined average values of $c$ for loops with high harmonics, but we did not take into account the fact that these loops would naturally self-intersect within a period of their evolution, thereby producing daughter loops. This is important, as the LIGO bounds are based on the assumption that the loops being considered are not self-intersecting. Earlier work to determine the number of final cusps per period emerging from loops with higher harmonics which then self-intersect can be found in Ref~\cite{Copi:2010jw}, where they used numerical simulations to characterise attractor non-self-intersecting loop shapes, beginning with initial loops containing M higher harmonic modes. They argued that such loops have on average $M^2$ cusps, and discovered that, on average, large loops will split into $3M$ stable loops within two oscillation periods (independently of M), with the stable loops being described by a degenerate kinky loop, co-planar and rectangular. These final loops were found to have a 40\% chance of containing a cusp. In this section, we will introduce a toy model that estimates $c$ based on a combination of analytical assumptions about the statistical properties of the loop self-intersection. 

Our motivation for developing this model is to estimate the contribution to the GWB signal from high harmonic cosmic string loops that chop off the long string network. These loops could potentially support a large number of cusps that could emit an enhanced GWB signal. Earlier approaches to calculate the gravitational wave signal from cosmic string loops have generally assumed that the cosmic string network follows the one scale model, and that loops do not undergo any further self-intersection, while at the same time it is assumed that loops contain roughly one cusp per period, $c=1$ \cite{Damour:2001bk, Abbott:2017mem, LIGOScientific:2021nrg}. While this may be a reasonable approximation, it does mean that a large signal that could be emitted from high harmonic cosmic string loops chopping from the long string network is ignored. Such a possibility was recently investigated by us in \cite{Pazouli:2020qmj} where we pointed out that such high harmonic loops can have many cusps and in principle can influence the overall GW signal considerably. However, we did not have a model of how the loops themselves could self-intersect and lead to a family of non-self-intersecting loops. 

The picture we have in mind is one in which at any point in the network evolution there is a significant number of large string loops (possibly containing high harmonics), which subsequently self-intersect forming daughter loops that can also self-intersect, and so on until non-intersecting descendant loops are formed. The scaling nature of the long-string network means that approximately half of the string network length gets transferred to newly formed loops in each Hubble time. While most of these loops will self-intersect within one oscillation period, the relevant timescale will be of order one tenth of a Hubble time ($\alpha\simeq 0.1$) and so such high-harmonic loops can provide a significant correction to the standard calculation of the gravitational wave signal from strings. In this context, we will aim to quantify the impact of multiple cusp events from high-harmonic order loops by modelling the loops' self-intersection history, thereby allowing us to provide an integrated effect for the value of $c$, based on cusps being formed at different stages of the loops' lifetimes, while at the same time keeping all the assumptions of the one scale model. 

To determine this integrated effect, there are a number of steps that we 
will follow. First we need to assume the distribution of harmonics on the
loops that have chopped off the long string network. For any given loop, 
we can calculate the number of cusps that it will produce during its lifetime, for a given probability that it might self-intersect, forming 
two smaller loops with each self-intersection. The chopping process is modeled using the binary tree evolution introduced in Section~\ref{assumptionstoymodel} below. At each level of the tree, the 
probability that a loop will chop is given by the results of Siemens and Kibble in Ref~\cite{Siemens:1994ir} (in which they determined the probability of an odd-harmonic string of harmonic order $N$ to 
self-intersect), while the number of cusps per period on the loop being analysed was obtained by us in \cite{Pazouli:2020qmj}. The number of cusps per period produced from each loop is thus averaged over the loop's
lifetime, by considering all the possible evolutions it might have (i.e. adding different binary trees, each with an assigned probability). Having calculated this quantity for odd-harmonic strings for several harmonic 
orders, we can compute the cusps per period for a given distribution of harmonics of the parent loops, which provides us with an estimate for $c$ for a network of string loops that chop off following such a binary tree 
evolution. We will begin by making the simplest assumptions concerning 
the evolution, which is what makes this a toy model. In Section~\ref{improvmodel}, we will discuss how these assumptions could be improved.

Note that there have been other publications where it was assumed that the evolution of a string can be imitated with a binary tree. In \cite{PhysRevD.36.987}, the binary tree had nodes corresponding to loops that chop with a given constant probability, that was independent of the harmonic order of the initial loop, or the tree level. Moreover the maximum number of tree levels was also not restricted by the harmonic order of the initial loop. The authors found that if the probability of self-intersection is larger than 1/2 there is a probability that the loops would chop infinitely, i.e. binary trees occurred with an infinite number of tree levels. A similar approach was assumed by Bennett in \cite{PhysRevD.33.872}, where the production and absorption of cosmic string loops from the long string network was studied, with smaller loops becoming less likely to re-absorb on the long string network as time went on. Bennett modeled the loop self-intersection by assuming that for each loop there was a probability to split into two equally sized daughter loops. He also assumed that the two daughter loops would oscillate with half the period of their parent loop and, therefore, they would split faster into two equally sized loops, compared to their parent loop. In \cite{Scherrer:1989ha}, Scherrer and Press tackled the problem of loop fragmentation in a numerical manner. Two different families of cosmic string loops were assumed, and each family of strings was tested for self-intersections, their evolution tracked using a numerical method until stable non-self intersecting loops were produced. They found that the probability of chopping was not a constant but it reduced with each loop generation. Moreover, the splitting of the loops did not necessarily occur in half, but over all length scales in a rather uniform distribution, i.e. the production of very small daughter loops was also observed. It was also found that if one assumes the splitting of loops into two with a probability that would decrease at each generation, then this analytic approach matched the numerical results well, when it comes to the daughter loops produced at each generation. We note with interest that they also determined that there was a correlation between the harmonic order of the parent loop, and the number of daughter loops, but it was not studied as to whether the maximum number of generations also depended on the harmonic order of the parent loop.

Finally, we should comment on the fact that the total number of daughter loops described using a binary tree model increases exponentially with the tree height. In particular, a fully expanded binary tree of height $n$, i.e. one where all of its nodes split, has $2^{n}$ daughter loops. By fully expanded binary tree we mean that any node at any level of the tree splits, until it reaches height $n$. In the above, we imply that a single loop is a binary tree of height 0. In any intermediate situation, i.e. at any tree where the nodes split or do not split with some probability, the increase will be exponential but with a basis less than 2. Therefore, it is reasonable to anticipate that any quantity that is a linear function of the number of daughter loops will inherit an exponential behavior. 
With our toy model we will provide an evaluation of the average cusp number produced per fundamental period $T=l/2$ from a cosmic string network of loops in a unit spacetime volume, $\nu(t)$. At the same time, we will not use the results of the toy model to make changes to any of the quantities of the ``one-scale" model of section \ref{CosmicStringNet}. Rather, our results will provide an integrated value for the parameter $c$, for a string network that otherwise follows the ``one-scale" approach. We will use the properties of the odd-harmonic family of cosmic string loops as provided in \cite{Siemens:1994ir} and \cite{Pazouli:2020qmj}, and modify the estimation of the cusps per period that we provided in \cite{Pazouli:2020qmj} by calculating an effective value for $c$ with the toy model.

A nice feature of our model is that it will also provide a means of estimating the number of stable non-self intersecting loops originating from a parent loop. By parent loop we define a cosmic string loop which has been produced from the long cosmic string network and has not yet self-intersected \citep{Copi:2010jw}. Note that in the ``one-scale" model of section \ref{CosmicStringNet}, all loops are parent loops since they do not self-intersect. However, here we will allow them to self-intersect and calculate an integrated value of $c$, over their lifetime.

In what follows, in Section~\ref{assumptionstoymodel}, we will begin by discussing the assumptions of the toy model. Then we will derive formulae for the stable loop number and the number of cusps produced by each loop. In order to carry out the analytic calculations we will use the symbolic software Mathematica, and will describe the details of the calculation in Appendix B, Section~\ref{ImplMath}. We will then calculate the cusps produced from a unit volume of the cosmic string network, $\nu(t)$, using the results of Section~\ref{CosmicStringNet}. Finally, in Section~\ref{improvmodel}, we will discuss the limitations and possible issues of this toy model, as well as ideas of how it could be improved.

\subsection{Assumptions of the model}\label{assumptionstoymodel}
First of all, we will assume that the loops maintain their Nambu-Goto nature and can be described via any type of Nambu-Goto loop solution. In \cite{Siemens:1994ir}, a specific Nambu-Goto loop set of solutions called the odd-harmonic string was presented and we tested their behavior at cusp points in \cite{Pazouli:2020qmj}. We will assume that at any stage of evolution the loops belong to this set of solutions. We will call the initial loop chopped from the long string network the parent loop, and all the loops produced through self-intersections starting from the parent loop, daughter loops. A loop will self-intersect if and only if the equation 
\begin{equation}
\vec{x}(\sigma,t)=\vec{x}(\sigma',t)
\label{selfinter}
\end{equation}
has at least one solution. In the above, $\sigma$ and $\sigma'$ both belong in the interval $[0,2\pi)$. After a self-intersection happens, the parent loop will split into two daughter loops which will have different initial conditions than the parent loop. If equation \eqref{selfinter} has no solution then the parent loop is a stable loop, i.e. it will evolve periodically without any self-intersection occurring in its lifetime. We will assume that each loop will self-intersect at one point only, producing two daughter loops, excluding the remote case of a loop self-intersecting at two points simultaneously. The daughter loops produced will be checked for self-intersections as we did in the case of the parent loop, and they may or may not produce more daughter loops. Note that equation \eqref{selfinter} was used to calculate the probability of self-intersection of an odd-harmonic loop in \cite{Siemens:1994ir}, and we will use these results to find the probability of a loop to split.

Eventually, the system will reach an equilibrium state where all possible self-intersections have happened, and it will consist of a number of stable loops, all produced from the initial parent loop. Note that we are sure that the equilibrium stage will be reached, since our assumed self-intersection probability of the loops (obtained from \cite{Siemens:1994ir}) is decreasing with the harmonic order, unlike in \cite{PhysRevD.36.987}, where under a given probability the system of loops can self-intersect indefinitely. At any stage of evolution, when we refer to the system at some time $t$ we will mean all the daughter loops produced from the initial parent loop that exist at that time, or in the case that the parent loop does not self-intersect the term system will refer simply to the parent loop.

Every time a loop self-intersects its length is reduced and divided between the two daughter loops. If $l$ is the length of the initial loop, and $l_1$ and $l_2$ the lengths of the daughter loops, then it does not necessarily hold that $l=l_1+l_2$, since some of the initial loop energy turns into kinetic energy of the daughter loops. However, we will assume that the above equality holds and the kinetic energy is negligible. The simplest scenario for how the length is divided is to assume that it is halved, i.e. that the daughter loops have equal length $l_1=l_2$. We will assume that this is the case for any self-intersection that occurs, whilst acknowledging that it is an important restriction we are imposing on the loop evolution. Since the length is halved, the fundamental harmonic of the loop that chops, which is the one with the longest wavelength, will no longer be present on the daughter loops, which will therefore have a total number of harmonics smaller than those of the parent loop. We will fix how the harmonics transition after a chopping, assuming that the daughter loops have harmonic order $N_i-2$ given that they were produced from a loop of harmonic order $N_i$. This ensures that the total harmonics are reduced as the loops chop, and they maintain the odd number of harmonics format. We will also assume that any self-intersection occurs after the loop has oscillated for half of its period. This choice is also based on the average value for the time it takes for a loop to self-intersect. In this toy model, there will be no need to determine the size of the loops, which is a subject of debate as we discussed in section \ref{CosmicStringNet}.

The splitting of the parent loop forms a full binary tree, which we will simply call binary tree or tree in the following \footnote{It is called tree because of its structure, full because each point (in our case loop) splits into a number from 0 to n daughter points, and binary because the number of possible daughter points is necessarily two in our case.}. We will call ``internal nodes'' the points (in our case loops) that are linked to points at the next level of the tree, and ``leaves'' the points of the tree that are not linked with points at the next level. The height of the tree $h$ is the number of levels it has. A tree that consists of a single leaf has a total height zero. The top level of the tree is the level at height zero, while the bottom level is the level corresponding to its maximum height. We will denote by $N$ the harmonic order of the highest harmonic order loop, which occurs at height $h=0$ and corresponds to the parent loop. Then, a loop of order $N_i$, which corresponds to one of the daughter loops after $i$ splittings of the parent loop, occurs at level of height 
\begin{equation}
h=\frac{N-N_i}{2},
\label{heightNi}
\end{equation}  
where $N$ and $N_i$ obtain odd integer values. The index $i$ corresponds to the height of the tree level. Each possible system which evolved from a parent loop, as we defined it above, can thus be described using a binary tree.  

We can assign a function $P(N_i,N_{i-2})$ between any two neighbouring levels of the tree, which is the probability of the loop of order $N_i$ splitting to two loops of order $N_{i-2}$. Also, we assume that for any loop, the left-moving and right-moving functions have an equal number of harmonics, as was assumed in \cite{Siemens:1994ir} and \cite{Pazouli:2020qmj}. The impact and significance of our assumptions on the model will be discussed in Section \ref{improvmodel}.

\subsection{The average number of stable loops and cusps produced from the parent loop}\label{stableloopcuspscalculation}
Below, we present our method for calculating the splitting of a parent loop, described by the odd-harmonic string, as well as the evaluation of the average number of stable loops and the average number of cusps emitted from the system over its lifetime, with respect to the harmonic order of the parent loop.

At each harmonic order we have a total number of possible binary trees, which is known and given below in equation \eqref{ctreenumber}, Section \ref{calculationanyharm}. To find the average number of stable loops from a parent loop of harmonic order $N$, we calculate the number of stable loops for each of the possible binary trees and then we average over all the binary trees, given that we know the probability of each binary tree to occur. The number of cusps is computed in a similar manner. We calculate for each possible binary tree the value of the total number of cusps  produced by the system divided by the total number of periods of the system. By total number of cusps produced we mean the sum of all cusp events that occurred in the system of loops until all loops in the system have vanished, i.e. in the lifetime of the system. By total number of periods we mean the number of periods of the parent loop (which is equal to $T_l=l/2$) that have occurred in the lifetime of the system (which is $\tau_l=l/\Gamma G\mu$). Note that the lifetime of the system is always less than or equal to the lifetime of the parent loop. Once more, averaging over the values for all binary trees we obtain the final result for the average number of cusps per period of an N harmonic order loop. We can then use this value of cusps per period, $c$, as an estimate for the number of cusps per period produced from an N order harmonic string throughout its lifetime, as part of a cosmic string network following the ``one-scale" evolution.

\subsection{Calculation for any harmonic order}\label{calculationanyharm}
For the probability of self-intersection of an $N$ order odd-harmonic loop, we will use the values obtained from Figure 3 in \cite{Siemens:1994ir}. In table \ref{tableselfinter}, we present these probability values in terms of the harmonic order. The probability value of an $N_i$ harmonic order loop not to self-intersect is $P(N_i,N_i)=1-P(N_i,N_i-2)$. The number $\varepsilon$  in Table \ref{tableselfinter} associated with $N_i > 11$ is considered to be much smaller than unity, which indicates the very low probability of a high-harmonic loop not to self-intersect. We will set it to be $\varepsilon=0.01$. The first order harmonic loop has zero chance of self-intersection, i.e. $P(1,1)=1$. 
\begin{table}[h]
\centering
\begin{tabular}{ |c||c|c|c|c|c|  }
 \hline
 Harmonic & Probability of  \\
 order $N_i$ & self-intersection\\
 \hline
 3       &  P(3,1)=0.6\\
 5       &  P(5,3)=0.8\\
 7       &  P(7,5)=0.9\\
 9       &  P(9,7)=1-0.05\\
 $\geq$ 11 &   P($N_i$,$N_i$-2)=1-$\varepsilon$\\

 \hline
\end{tabular}
\caption{The probability of a loop of harmonic order $N_i$ self-intersecting into two loops of harmonic order $N_i-2$, taken from reference \cite{Siemens:1994ir}.}
\label{tableselfinter}
\end{table}

For the number of cusps per period of a loop of a given harmonic order $N_i$, $c_{N_i}$, we will use the values obtained for the odd-harmonic loops presented by us in \cite{Pazouli:2020qmj}. Regarding the total number of periods, in terms of the period of the parent loop, this will be calculated by dividing the total lifetime of the tree by the period of the parent loop. The lifetime of stable loops is significantly larger than the lifetime of a loop that self-intersects, since $\Gamma G \mu \leq 10^{-6}$ and so, while the period is $l/2$, the lifetime of a non-self interesting loop is $l/\Gamma G\mu$ (see III.B). Also, the lifetime of a stable loop is larger the smaller its tree height. Therefore, the total lifetime of the tree is given by the sum of the lifetime of the stable loop with the smallest tree height plus the lifetime of the loops that precede it.  

Let us denote by $a(h)$ the number of trees of a given height $h$, where $h \geq 0$ takes positive integer values. Then, it can be proven that the recurrence relation \cite{OEISah2020}
\begin{equation}
a(h+1)=a(h)^2+2a(h)\left[a(h-1)+a(h-2)+\dots+a_0\right],\quad a_0=1
\label{treenumber}
\end{equation}  
holds.
If we also denote by $b(h)$ the cumulative number of trees up to height $h$ we know that it is also expressed by the following recurrence relation \cite{OEISbh2020}
\begin{equation}
b(h+1)=b(h)^2+1,\quad b_0=1.
\label{ctreenumber}
\end{equation} 
Note that the zero height values are obvious, since they correspond to a single point. The simplicity of relation \eqref{ctreenumber}, allows us to find the total number of all trees with height from 0 to $h$. This describes the total number of configurations we can potentially have when a loop of harmonic order $N$ self-intersects, corresponding to trees of maximum height $h=(N-1)/2$. We notice that the number of trees increases in a recurrence power law manner. This implies that the value for height $h$ is the square of the previous value, which is the square of the value before that, and so on and so forth, corresponding to a very rapid increase of the total cumulative number of trees with respect to height. 

We will now define the trees with same type, i.e.\ the trees that have the same number of leaves and internal nodes at each level. Note that this definition implies that these trees have the same height, $h$ too. An example of trees of the same type are the trees that appear in figures \ref{5harmonicevolution}(c) and \ref{5harmonicevolution}(d). We define the multiplicity (also called cardinality) $d(h)$, as the number of different types of trees with height $h$. For example, from figure \ref{5harmonicevolution} we can see that $d(2)=2$. The recurrence series of the tree multiplicity is given in \cite{OEISdh2020}. We can also define the degeneracy of a tree type $D(h,i)$, to be the number of trees of height $h$ that belong to the same tree type. The index $i$, running from $1$ to $d(h)$, labels the different tree types at height $h$. For example, from figure \ref{5harmonicevolution}(c) and \ref{5harmonicevolution}(d), it follows that for $h=2$ the degeneracy is $D(2,1)=2$, since we have two degenerate loops in that tree type. Given the above, we can also write the number of trees of a given height in terms of the tree degeneracy
\begin{equation}
a(h)=\sum_{i=1}^{d(h)} D(h,i)
\label{treenumber2}
\end{equation} 
and the cumulative number of trees up to height $h$ is written as
\begin{equation}
b(h)=\sum_{j=0}^{h}a(j)=\sum_{j=0}^{h}\sum_{i=1}^{d  (j)} D(j,i).
\label{ctreenumber2}
\end{equation}

\subsubsection{Calculation of the average number of stable loops}\label{asloops}
In our model we deal with loops of harmonic order $K$, where $K$ is odd. Each loop can self-intersect producing other (lower) odd-harmonic order loops. This process can continue and loops of harmonic orders $K-M$ are produced, where $M$ takes even values and satisfies $0\leq M \leq K-1$. Then, the tree height is given by $h=M/2$.

We will define the final harmonic order of such an evolving string loop to be the smallest harmonic order of any of the stable loops of the system, $K-M$. Note that this does not prevent the system to also include stable loops with harmonic order greater than $K-M$. 

Let us now derive a formula for the average number of stable loops formed from a parent loop of harmonic order $K$. We should start with the calculation of the number of stable loops for trees with fixed tree height, between harmonic order $K$ and $K-M$. This is described by the quantity 
\begin{equation}
f_{K,K-M}=\sum_{i=1}^{d_{K,K-M}}c_{K,K-M}^i P^i(K,K-M) D_{K,K-M}^i,
\label{averstloseries}
\end{equation} 
and includes all trees of fixed height $M/2$. In the above, $d_{K,K-M}$ is $d(M/2)$, which is the number of different types of trees with height $M/2$. The quantity $c_{K,K-M}^i$ is the number of stable loops (leaves) of trees with height $M/2$ and of the same type $i$. Also, $P^i(K,K-M)$ is the probability of trees of type $i$ and height $M/2$ to occur. As mentioned above, each tree configuration has a given probability of occurring, and the total probability of trees of type $i$ is the sum of the probabilities of all trees of type $i$. Finally, the quantity $D_{K,K-M}^i$ is the degeneracy of trees of the same type $i$ and height $M/2$. Note that we use the notation $d_{K,K-M}$ and $D_{K,K-M}^i$, instead of $d(M/2)$ and $D(h=M/2,i)$, respectively, to specify the initial and final harmonic orders for the calculation in our summation formula \cite{Pazouli:2020qfr}.

Given the above, we can calculate the average number of stable loops produced from a parent loop of harmonic order $N$ if we sum the above quantity $f_{K,K-M}$, defined in \eqref{averstloseries}, over all possible tree heights, i.e. ranging from $0$ (which corresponds to the case where the parent loop does not self-intersect) to $(N-1)/2$ (the maximum tree height for an $N$ order harmonic loop that is allowed to self-intersect down to first order harmonic loops). 

If we denote the average number of stable loops of an $N$ order harmonic string by $sl_{N}$, we can thus write
\begin{equation}
sl_N=\sum_{i=0}^{\frac{N-1}{2}}f_{N,N-2i},
\label{slm}
\end{equation} 
where $N$ is odd. Note that the subscript where $N$ appears twice (i.e. subscript $N,N$ for $i=0$) implies an $N$ harmonic order loop that does not self-intersect. Since the minimum harmonic order is $1=N-(N-1)$, the maximum height is $(N-1)/2$. The summation of all tree configuration probabilities over all possible tree heights is equal to 1 by definition (see section \ref{ImplMath}).

\subsubsection{Calculation of the average number of cusps per period}
For the calculation of the average cusp number, we will need to define both the period of the daughter loops, and the total number of periods in the lifetime of the system (see section \ref{stableloopcuspscalculation}). As we discussed in section \ref{assumptionstoymodel}, in our toy model a loop that self-intersects will split into two equal sized loops, and since the period of a loop is proportional to its length, this implies that the period of the two new loops will be half that of their parent loop. Therefore, we can write a recursion relation for the periods of a loop of harmonic order $K-M$ emerging from a loop of harmonic order $K-M+2$
\begin{equation}
T_{K-M}^{(K)}=\frac{T_{K-M+2}}{2}
\label{periodkl}
\end{equation}  
where $2\leq M\leq K-1$. The superscript $(K)$ indicates the harmonic order. It is also useful to know how long a loop lives. If the loop self-intersects, it will live for half its period, according to our assumptions in section \ref{assumptionstoymodel}. Therefore, the expression for its lifetime is 
\begin{equation}
\mathcal{T}_{K-M+2,K-M}^{(K)}=\frac{T_{K-M+2}^{(K)}}{2},
\label{lifetimekl2}
\end{equation}  
where $2\leq M\leq K-1$. The notation $\mathcal{T}_{K-M+2,K-M}^{(K)}$ means the lifetime of a loop of harmonic order $K-M$ created from a loop of harmonic order $K-M+2$, with parent loop of harmonic order $K$. If the loop does not self-intersect, then if formed at time $t_i$ its lifetime (see  \eqref{lengthtime}) is given by  
\begin{equation}
t=\frac{(\alpha+\Gamma G\mu) t_{i}}{\Gamma G\mu}.
\label{looplifetime}
\end{equation}

In this case, the lifetime of the loop (for $\alpha\gg \Gamma G\mu$) is 
\begin{equation}
\mathcal{T}_{K-M,K-M}^{(K)}=\frac{T_{K-M+2}^{(K)}}{\Gamma G\mu}.
\label{lifetimekl}
\end{equation}
We remind the reader of the notation being used in \eqref{lifetimekl2} and \eqref{lifetimekl}, the subscript $K-M,K-M$ indicates a loop of order $K-M$ that does not self-intersect, 
while the subscript $K-M+2,K-M$ indicates a loop of order $K-M+2$ that self-intersects, splitting into two $K-M$ loops.  

The total lifetime of the system will be the total time from the moment the parent loop of harmonic order $K$ is created until all of the loops that were created via self-intersections have evaporated. Note that the lifetime of a stable loop is significantly larger than a loop that self-intersects, because $\Gamma G\mu \leq 10^{-6}$. From equation \eqref{lifetimekl}, we can see that the larger the harmonic order $K-M$ the longer the stable loop lives. Therefore, the total lifetime of a system with leaves having maximum harmonic order $K-M$ 
will be given by
\begin{equation}
\mathcal{T}_{K-M}^{(K)}=\mathcal{T}_{K,K-2}^{K}+\mathcal{T}_{K-2,K-4}^{K}+\dots +\mathcal{T}_{K-M-2,K-M}^{(K)}+T_{K-M,K-M}^{(K)}.
\label{lifetimetotal}
\end{equation}
The highest order harmonic leaf does not necessarily correspond to a single loop of the system, since there can be multiple stable loops of the same harmonic order. Equation \eqref{lifetimetotal} summarizes the lifetimes of loops from harmonic order $K$ to harmonic order $K-M$, which corresponds to a stable loop.

For a parent loop with harmonic order $K$ formed at time $t_i$ we need $M/2$ steps in the tree (i.e. $M/2$ differences in height) to reach a loop of harmonic order $K-M$. Therefore, the period of the $K-M$ harmonic order loop can be written as
\begin{equation}
T_{K-M}^{(K)}=2^{-\frac{M}{2}}T_{K}^{(K)}=2^{-\frac{M}{2}}\frac{\alpha t_i}{2}=2^{-\frac{M+2}{2}}\alpha t_i,
\label{periodrec}
\end{equation}
where we have used the fact that the period of the loops is halved at each intersection. The harmonic order $M$ takes integer values in the interval $[0,K-1]$. The lifetime of a loop of harmonic order $K-M+2$ which splits (into a $K-M$ harmonic order loop) is
\begin{equation}
\mathcal{T}^{(K)}_{K-M+2,K-M}=\frac{T_{K-M+2}^{(K)}}{2}= T_{K-M}^{(K)},
\label{lifekl2rec}
\end{equation}
and we can write it in terms of the period of the parent $K$ harmonic loop as
\begin{equation}
\mathcal{T}^{(K)}_{K-M+2,K-M}=2^{-\frac{M+2}{2}}\alpha t_i
\label{lifekl2simple}
\end{equation}
where $2\leq M \leq K-1$. If we apply the parameter transformation $M\rightarrow M+2$, we find that equation \eqref{lifekl2simple} is written as
\begin{equation}
\mathcal{T}^{(K)}_{K-M,K-M-2}=2^{-\frac{M+4}{2}}\alpha t_i,
\label{lifekl2simple2}
\end{equation}
where $0\leq M\leq K-3$. Combining equations \eqref{periodrec} and \eqref{lifekl2simple2}, we find that,
\begin{equation}
\frac{\mathcal{T}_{K-M,K-M-2}^{(K)}}{T_{K-M}^{(K)}}=\frac{1}{2}
\label{lifetoperiodnode}
\end{equation}
with $0\leq M\leq K-3$. For the case of a stable loop, combining equations \eqref{periodkl} and \eqref{lifetimekl}, we find that
\begin{equation}
\frac{\mathcal{T}_{K-M,K-M}^{(K)}}{T_{K-M}^{(K)}}=\frac{T_{K-M+2}^{(K)}}{\Gamma G\mu T_{K-M}^{(K)}}=\frac{2 T_{K-M}^{(K)}}{\Gamma G\mu T_{K-M}^{(K)}}=\frac{2}{\Gamma G\mu}.
\label{lifetoperiodst}
\end{equation}

Given the above results, we can calculate the total number of cusp events produced by a tree with starting harmonic order $K$ and finishing with order $K-M$
\begin{equation}
g_{K,K-M}=\sum_{i=1}^{d_{K,K-M}} P^i(K,K-M) \tilde{c}_{K,K-M}^i D_{K,K-M}^i,
\label{gkl}
\end{equation}
using the same reasoning as in equation \eqref{averstloseries}. The upper sum limit $d_{K,K-M}$ is $d(M/2)$, i.e. the number of different types of tree with height $M/2$. The probability $P^i(K,K-M)$ is the probability of trees of type $i$ and height $M/2$. Also, $D_{K,K-M}^i$ is the degeneracy of trees of the same type $i$ and with height $M/2$. Finally, the quantity $\tilde{c}_{K,K-M}^i$ corresponds to the total number of cusps produced from the nodes and the leaves of the tree of type $i$
\begin{equation}
\begin{aligned}
\tilde{c}_{K,K-M}^i=&\sum_{j=0}^{M/2-1}\frac{\mathcal{T}^{(K)}_{K-2j,K-2j-2}}{T^{(K)}_{K-2j}}c_{K-2j}n^i_{(K,K-M,K-2j)}\\
&+\sum_{j=0}^{M/2}\frac{\mathcal{T}^{(K)}_{K-2j,K-2j}}{T^{(K)}_{K-2j}}c_{K-2j}l^i_{(K,K-M,K-2j)}.
\label{ctildekl}
\end{aligned}
\end{equation}
Note that by total number of cusps we mean all the cusp events produced from the creation until the evaporation of the loop of type $i$ and height $M/2$. In \eqref{ctildekl}, the first sum summarizes the cusp contribution from the nodes (i.e. loops that split) of the tree and $n^i_{(K,K-M,K-2j)}$ is the number of nodes at tree height $K-2j$ for a loop with initial harmonic order $K$ and final $K-M$. The second sum in \eqref{ctildekl} summarizes the cusp contribution from the leaves (i.e. stable loops) of the tree and $l^i_{(K,K-M,K-2j)}$ is the number of leaves at tree height $K-2j$ for a loop with initial harmonic order $K$ and final $K-M$ \cite{Pazouli:2020qfr}. The nodes contribute cusp events for the half of their period, $\mathcal{T}^{(K)}_{K-2j,K-2j-2}$, after which they split. The leaves contribute for much longer since they do not split, $\mathcal{T}^{(K)}_{K-2j,K-2j}$. The parameter $c_{K-2j}$ is the number of cusps per period of an odd-harmonic loop of harmonic order $K-2j$, the values of which are given in Table 1 of \cite{Pazouli:2020qmj}. The fractions in equation \eqref{ctildekl} give the number of periods for which the corresponding loop oscillates until it chops or evaporates.

Having calculated the total number of cusps produced for a tree configuration of a given height, we will sum over the possible tree heights. We will also normalize the result with respect to the total number of periods the $K$-th order harmonic parent loop oscillates, $2/\Gamma G\mu$, to finally obtain the average cusps per period of a $K$-th order harmonic loop 
\begin{equation}
c^{(K)}=\frac{\Gamma G\mu}{2}\sum_{i=0}^{\frac{K-1}{2}}g_{K,K-2i}.
\label{cfinalk}
\end{equation}
In deriving the above equation, we followed the same reasoning as in our derivation of the stable loop number formula \eqref{slm} in section \ref{asloops}.

\subsection{Number of stable loops and cusps per period with harmonic order}
The number of different tree configurations increases rapidly with the tree height. As we can see from the recurrence relation \eqref{ctreenumber2}, the number of configurations is 1 for $h=0$, 2 for $h=1$, 5 for $h=2$, 26 for $h=3$, 677 for $h=4$, and so on so forth. In Appendix \ref{lowharmtree} we show how the binary tree builds up for harmonics from 
$N=1$ to $N=5$, but it is clear these numbers rise rapidly and require a numerical algorithm if we are to determine the number of stable loops and cusps per period for $h \gg 2$. We present such a method in Appendix \ref{ImplMath}, and also calculate the results numerically using a Monte Carlo approach, to compare with our analytic results. In this section we present and compare our results for the number of stable loops and cusps per period as a function of harmonic order.

Specifically, in Table \ref{tableIM}, we present the results that were obtained using the analytic method that was implemented with Mathematica. This method allows us to reach results for harmonic order of the parent loop ranging from 1 to 17, the limit coming from the fact that the computational time increases rapidly with harmonic order. The number of cusps per period at harmonic order 17 was not possible to be calculated due to the computational time required. We find that the Monte Carlo method is much faster than the analytic one and allows us to calculate data for higher harmonic loops. The results of both approaches are the same up to the third significant figure. 

It is worth comparing the data for the average number of cusps per period from the stable loops in Table \ref{tableIM} with the number of cusps per period analysed in \cite{Pazouli:2020qmj} for a class of odd-harmonic-loops which did not take into account any self-intersections. For harmonics below $N=5$, the two are comparable, but above that the toy model numbers increase dramatically compared to \cite{Pazouli:2020qmj}, increasing exponentially rather than linearly. In fact using the least squares method to the data from Table \ref{tableIM}, we find that the number of cusps are related to the harmonic order via 
\begin{equation}
\ln (c)=a+ \ln(\beta)N\Rightarrow c=e^a \beta^{N}
\label{cexpN}
\end{equation}
where $a=0.348$ and $\ln (\beta)=1.42$. These exponentially large cusp numbers for high harmonic loops should be taken with a grain of salt given the ansatz we are making about the self intersections of a loop into two equal sized loops. This is too prescriptive, especially for the high harmonic cases where it is likely much smaller loops will be chopped off the network first. With that in mind, we focus our attention on the results for the lower harmonic parent loops from $N=1$ to $N=7$. We discuss this issue further in section \ref{improvmodel}.

\begin{figure}%
\centering
\includegraphics[scale=0.6]{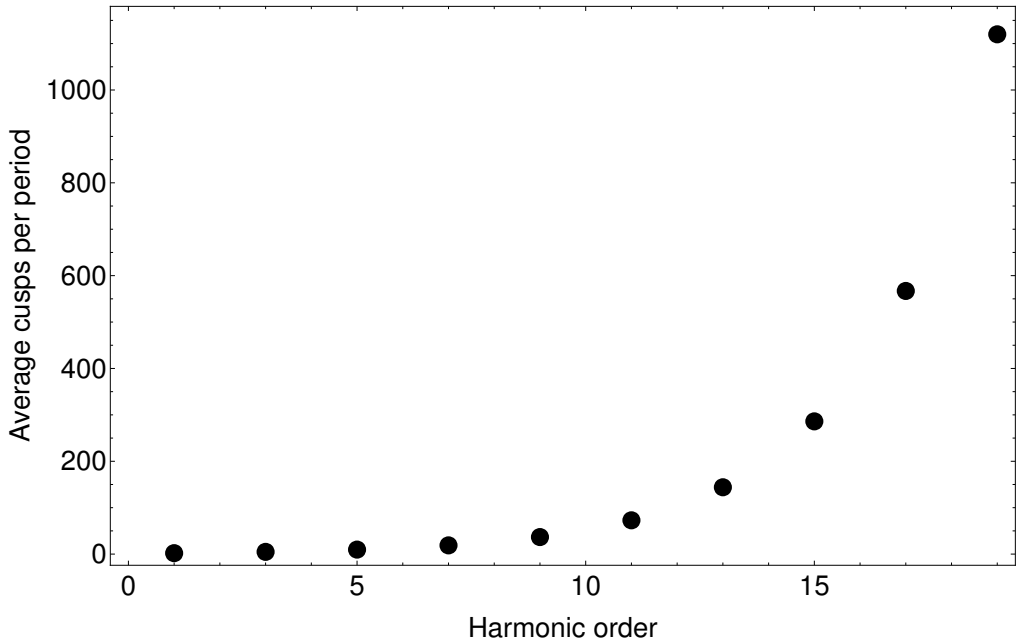} %
\caption{The average number of cusps per period vs the harmonic order of the parent loop.}%
\label{cMCplot}%
\end{figure}

\begin{table}
\centering
\begin{tabular}{ |c||c|c|  }
 \hline
 Harmonic & average number of  & average number of cusps \\
 order &   stable loops & per period \\
 \hline
 1 &  1 &  2.00 \\
 3 &  1.6 & 4.8 \\
 5 &  2.76 & 9.68 \\
 7 &  5.07 & 18.8 \\
 9 &  9.68 & 36.7 \\
11 &  19.2 & 72.8 \\
13 &  38.0 & 144 \\
15 &  75.2 & 286 \\
17 & 154 & Unknown \\
 \hline
\end{tabular}
\caption{The average number of stable loops and cusps per period calculated using the method described in section \ref{ImplMath}, for loops of harmonic order 1 to 17. Note that the unknown value at harmonic order 17 is due to very long execution time.}
\label{tableIM}
\end{table}

\subsection{Calculation of the number of cusps per period for a cosmic string network}
\label{Cuspsperperiodnetwork}
When we calculate signals from a network of cosmic strings, we will need some additional assumptions regarding the harmonic order distribution of the strings in the network to implement our results. In particular, regarding the gravitational wave signal, the number of cusp events per unit spacetime volume is \cite{Damour:2001bk}
\begin{equation}
\nu(t)=\frac{cn_{l}(t)}{T_{l}},
\label{cuspsdv}
\end{equation}
where $c$ is the average number of cusps per period, $T_{l}=l/2=\alpha t$, and $n_{l}(t)$ is the number density of loops. The value of $c$ will be calculated in this section, using equation \eqref{cfinalk} and an assumed distribution for the harmonics of the loops in the string network. It is usually assumed that $c \sim 1$. To compute the value of $c$ using our results from the toy model in Section~\ref{toymodelevol}, we will need to assume a distribution of the harmonic order of the loops in a unit volume. There is not a great deal known about this, so we will follow a conservative approach and assume that low harmonic loops dominate over high harmonic loops when created from the long string network, and also that the density of loops of a given harmonic order drops with the harmonic order. Further, as we have just discussed, we will also take into account only loops with harmonic order from $N=1$ to $N=7$. 

Given these assumptions, we will aim to split the harmonic order distribution of loops in a cosmic string network volume into first, third, fifth and seventh order harmonic loops, following a discrete distribution. The simplest and most straight forward way to achieve this is to assume a uniform distribution of the aforementioned harmonics. In this case, the average cusp number per period from a unit spacetime volume will be the average of the values for the cusps per period presented in Table \ref{tableIM}, for harmonic order of the parent loop from $N=1$ to $N=7$. Denoting this as $c$, we find that it is given by $c=8.82$. We will use this result in section \ref{GWBcs}, to modify the current assumption for the cusps per period when estimating the amplitude of the gravitational waves originating from cusps on cosmic strings. The current assumption for the cusps per period from a unit volume is taken to be between $c=1$ and $c=0.1$ \cite{Damour:2001bk,Abbott:2017mem}. 

Although we have just presented one estimate for $c$ based on a uniform distribution of harmonics, the fact that we do not really understand the way the loops are distributed in terms of harmonic number suggests we should consider a number of possible cases. We do this below, with the only requirement for the discrete harmonic distribution being that it should drop off quickly with the harmonic order. A discrete distribution that satisfies the above requirement is Benford's law. A set of numbers $P(d)$ (in our case the percentage of loops of a given harmonic) given by Benford's law satisfy 
\begin{equation}
P(d)=\log_{b}\left(1+ \frac{1}{d}\right).
\label{}
\end{equation}
The parameter $b$ will be fixed to $b=5$, since we are interested in taking into account the harmonics $N=1$ to $N=7$ in the string network. This yields $P(1)=0.43$, $P(2)=0.25$, $P(3)=0.18$ and $P(4)=0.14$, which satisfy 
\begin{equation}
\sum_{d=1}^{4}P(d)=1,
\label{}
\end{equation}
as expected. Then, we find that for this distribution of the parent loop harmonics the average number of cusps per period from a unit volume is $c=6.43$. Note that this distribution provides us with only one choice of values for the distribution of harmonics of the parent loops.

Another distribution which provides us with possible values of the parent loop harmonics (that decrease as the harmonic order increases) is the geometric distribution 
\begin{equation}
G(p,k)=\left(1-p\right)^k p
\label{}
\end{equation}
where $k=\{0,\mathbb{Z}^{+}\}$ and $0<p\leq 1$. In the above, we can fix the value of $p$ and obtain the values $G(p,k)$ for harmonics $1$ to infinity, since $k$ can obtain the value of any natural number, with 
\begin{equation}
\sum_{k=0}^{\infty}G(p,k)=1.
\label{}
\end{equation}
This distribution has the advantage that it allows for different initial values of the density of parent loops of a given harmonic, by using different values for $p$, unlike the even distribution and Benford's law. However, since we would like to focus on loops of harmonics up to $N=7$, we will choose values of $p$, such that higher order parent loops are scarce in the unit volume. To choose such values, let us first produce the formula that calculates $c$ for this distribution. This is given by
\begin{equation}
c=\sum_{k=0}^{\infty}G(p,k)c_{2k+1}
\label{}
\end{equation}
where the subscript $2k+1$ corresponds to the harmonic order of the parent loop, which contains odd values only, and the values of $c_{2k+1}$ are given in Table \ref{tableIM}. Using equation \eqref{cexpN}, we find that 
\begin{equation}
c=\sum_{k=0}^{\infty}\left( 1-p\right) ^k p e^a \beta^{2k+1}.
\label{sumbarcg}
\end{equation}
By summing the above, we find that
\begin{equation}
c=pe^a \beta \frac{1}{1+(p-1)\beta^2},
\label{}
\end{equation}
for $(1-p)\beta^2<1$. Note that the sum is of the form of a geometric series, i.e. the sum of numbers in a geometric progression $x^k$, with $x=(1-p)\beta^2$. This sum is known to converge if and only if the above inequality holds. Since the value of $\beta$ has been found using the least square method, we find that the above sum converges if and only if $p>0.94$. For example, for $p=0.95$, we find that $c=38.9$. However, we are interested in the values of the sum from $k=0$ (which corresponds to $N=1$) to $k=3$ (N=7), and we would like to consider the case where the higher harmonic contributions are negligible by minimizing their appearance in the distribution. Therefore, regardless of the sum convergence, we can examine the behavior of the truncated sum for any value of $p$. We find that the values of $c$ with regards to $p$, calculated using the sum of the first 4 terms of \eqref{sumbarcg}, lie in the interval $[0,5.03]$. The maximum value of the plot corresponds to $p=0.35$, which has a relatively large percentage of high harmonic loops $\sum_{k=4}^{\infty} G(0.35,k)=0.18$, which are neglected. Preferably, we would like a value of $p$ that contains mainly lower harmonic order loops (up to $N=7$), such that it follows the pattern that the harmonic order loop distribution drops quickly with $N$. For $p=0.6$, the percentages of the parent loops of each harmonic order are given by, $G(0.6,0)=0.6$, $G(0.6,1)=0.24$, $G(0.6,2)=0.1$, $G(0.6,3)=0.04$. Finally, the rest of the harmonic order loops sum to a total percentage of $\sum_{k=4}^{\infty} G(0.35,k>3)=0.02$, which we will consider negligible. For this case, the value of cusps per period is $c=4.0$. Note that around this value of $p$, the value of $c$ is relatively stable, and the aforementioned value represents the distribution well. 

Given the above calculations, the three different distributions we used (the uniform distribution, Benford's law and the geometric distribution) provide a value of $c$ evaluated over the lowest four harmonics that does not vary significantly from distribution to distribution. In the case of the uniform distribution the harmonic order of the loops is equally distributed, while in the other two cases the percentage of loops of a given harmonic drops as the harmonic order increases. Based on the intuition obtained from numerical simulations of loops, we are disregarding cases where the percentage of loops increases as their harmonic order increases, which needs to be borne in mind when considering these results. Given the three estimates for $c$, namely $c \sim 8.82, 6.64$ and $4.0$ for the uniform, Benford and geometric distributions respectively, we see that for the case of harmonics up to $N\sim 7$, our approach suggests the average number of cusps on a non-self intersecting loop is typically of order $4-10$ suggesting it could have an impact on the results presented in \cite{Abbott:2017mem,LIGOScientific:2021nrg} where $c\sim 1$ was assumed. We will develop this argument in Section~\ref{results}.

\subsection{Possible improvements of the model}\label{improvmodel}
A possible criticism of the toy model is that kinks formed during the self-intersection process of the loops are ignored. As discussed in \cite{Copi:2010jw},\cite{vilenkin1994cosmic}, the occurrence of cusps is suppressed with the presence of kinks on the Kibble-Turok sphere. Therefore, this toy model gives an enhanced number of cusps per period output, which also rises rapidly as the harmonic order of the loops increases. A way to counter this enhancement of cusp events would be to add a suppression factor which would account for the kink effect on the cusp production, see for example \cite{Copi:2010jw}. Another factor that suppresses cusps is the gravitational back-reaction around cusps and kinks, which rounds off kinks and makes cusps weaker \cite{Blanco-Pillado:2018ael}, \cite{Blanco-Pillado:2015ana}. This phenomenon could also introduce a suppression factor for the cusp occurrence on the evolution of the loops. However, it could have a counter effect towards enhancement of the cusp number due to the rounding of the kinks. 

Another issue is the percentage of the lifetime of the $N_i$ harmonic self-intersecting loop, which we assumed to be half of the period. Could a better assumption work? An idea would be to check how many solutions equation \eqref{selfinter} has, i.e. how many possible self-intersections could happen, and split the lifetime to that fraction of the period. Also, the assumption of the harmonic order of the daughter loops always following the rule of being minus 2 of the loop that chopped would change in this picture, which would allow the loop to chop into smaller loops from earlier stages of evolution. This would also affect our assumption on the length of the daughter loops. For a numerical simulation of a loop chopping see \cite{Copi:2010jw,Scherrer:1989ha,Blanco-Pillado:2019tbi}. It was their numerical results that lead to our decision to only consider loops up to harmonic order $N=7$ of the parent loop. It appears likely that at higher harmonics a parent loop would quickly self-intersect into much smaller loops, and hence it would not follow our ansatz rule that harmonics decrease by 2 at each step. This would lead to a significant change in the values of stable loops and cusps that would be calculated.

The restriction of the odd-harmonics only loops implies that the strings produced cannot obtain any even harmonic values, which fixes the evolution of the loops to a particular shape. Given the freedom of any harmonic order, the binary trees would have different branches. Furthermore, the results we find could be heavily dependent on the string solution family that one assumes. For example, in \cite{Scherrer:1989ha}, two relatively similar families of stings are assumed, yet the difference in the results is significant. 

Also, recall that we do not take into account the kinetic energy lost every time a loop self-intersects and forms two daughter loops. If this energy could be a significant percentage of the system energy, it could cause the system to diminish faster.

Furthermore, the main essence of this model, which renders it analytical, is the stochastic behavior of the system of loops and their odd-harmonic string behavior at any stage of the evolution. This kind of assumption could be quite restrictive for such a complicated system and it could prove inadequate compared to a potential numerical simulation that follows the exact motion of the strings on a grid at every time interval of the system's evolution. Such models have been developed (see for example \cite{Scherrer:1989ha}) but they are not directly comparable to our model due to the different string configuration assumed. However, a numerical model developed using the odd-harmonic string and calculating the stable loops and the total cusps produced would be comparable with this toy model, and it could be a way of testing our assumptions and results.

Finally, our assumed harmonic distribution in the loops chopped-off the long string network in section \ref{Cuspsperperiodnetwork}, is somewhat restricted due to the limitations of our model, namely that the maximum harmonic of the loops chopped is $N=7$, and the fact that we assumed that the low harmonics would dominate, aiming to adopt a conservative approach. 

The aim of this toy model is to give a general idea of how the value of $c$, the number cusps per period from a spacetime volume of loops, could be enhanced if we took into account the high-harmonic loop contribution to cusp production in the ``one-scale" string network, and to probe the stochastic evolution of loops. Although the result of $c\sim 4 - 10$ appears to be fairly robust, in view of the three distributions we have used to model the harmonic content of parent loops in the network, we appreciate that we have had to made a number of assumptions that make this a toy model.

\section{Propagation of GWBs in FLRW spacetime}\label{PropGWBFLRW}
In this section we return to the cosmology, to estimate the GWBs emitted from Nambu-Goto closed cosmic strings, given our new estimates for the parameter $c$. In this approach the string is a one-dimensional object and its world history can be represented by a two-dimensional surface in spacetime, the worldsheet, which is described by the mapping functions
\begin{equation}
X^\mu=X^\mu(\tau,\sigma).
\label{}
\end{equation}
These map the worldsheet parameters ($\tau,\, \sigma$), used to describe the two-dimensional surface, to spacetime coordinates. The parameter $\sigma$ corresponds to the position on the string and it is subject to periodic identification, since we are considering closed loops of string. The strings follow a two-dimensional wave equation of motion in flat spacetime (we use the convention for the Minkowsi metric $\eta_{\mu\nu}=diag(-1,+1,+1,+1)$). We fix the gauge-invariance of the cosmic string solutions using the conformal gauge and the time gauge \cite{vilenkin1994cosmic}. The loops move periodically with fundamental period $T=l/2$, where $l$ is its invariant length, and frequency $\omega_1=4\pi/l$. The loop will emit gravitational waves at the discrete frequencies $\omega_m=m\omega_1 $, where $m$ belongs to the set of natural numbers and runs from 1, the fundamental harmonic order of the loop, up to the maximum harmonic order of the loop $N$. The movement of the loops is described as the sum of a left-moving wave $\vec{b}$ and a right-moving wave $\vec{a}$,
\begin{equation}
\vec{X}(t, \sigma)=\frac{1}{2}\left[\vec{a}(\sigma-t)+\vec{b}(\sigma+t)\right],
\label{lrm}
\end{equation}
which we will also refer to as the string movers. We will denote $u=\sigma-\tau$ and $v=\sigma+\tau$. Note that we have normalized the left- and right-movers by $l/2\pi$ and the dimensionless parameters $u$ and $v$ range in the interval $[0,2\pi)$, as in \cite{Siemens:1994ir,Pazouli:2020qmj}.

It was calculated in \cite{Damour:2001bk} that the GWB amplitude propagated in an FLRW flat spacetime, which is observed at redshift $z$, distance $r(z)$ and frequency $f=f_{rec}$ is given by 
\begin{equation}
\tilde{h}^{cusp}(f,z)=Cg_1\frac{G\mu l^{2/3}}{(1+z)^{1/3}f^{1/3}r(z)},
\label{hcuspprop1}
\end{equation}
where
\begin{equation}
g_1=\left(|\vec{a}''(u_c,v_c)||\vec{b}''(u_c,v_c)|\right)^{-1/3},
\label{g1def}
\end{equation}
and the dimensionless parameter $g_1$ is evaluated at the cusp point $(u_c,v_c)$,  $\vec{a}''(u_c,v_c) \equiv \frac{\partial^2 \vec{a}}{\partial (u)^2}$, etc. Also, $C$ is a constant, calculated to be $C/2\pi\simeq 0.8507$ \cite{Damour:2001bk,Pazouli:2020qfr}. The quantity $\tilde{h}$ is the logarithmic Fourier transform of the GWB amplitude \footnote{The logarithmic Fourier transform is defined as $\tilde{g}(k)=|k|g(k)=|k|\int_{-\infty}^{\infty} dt g(t)e^{-2\pi ik  t}$, which provides the advantage that $f(k)$ has the same physical dimensions as $f(t)$.}. The Fourier transform of the GWB amplitude is
\begin{equation}
h^{cusp}(f,z)=Cg_1\frac{G\mu l^{2/3}}{(1+z)^{1/3}f^{4/3}r(z)},
\label{hcuspprop2}
\end{equation}
which coincides with the result in \cite{Abbott:2017mem}. 

We can use the cosmic distance approximation, with $t_0$ denoting the age of the universe,
\begin{equation}
r(z)=\frac{t_0 z}{1+z}
\label{}
\end{equation}
provided in \cite{Damour:2001bk} to simplify the expression for the GWB amplitude, leading to 
\begin{equation}
\tilde{h}^{cusp}(f,z)= Cg_1\frac{G\mu l^{2/3}}{(1+z)^{1/3}f^{1/3}}\frac{1+z}{t_0 z},
\label{hcuspprop3}
\end{equation}
which is the result obtained in \cite{Damour:2001bk} and used in \cite{Abbott:2017mem}.

The value of the angle $\theta^{div}$, which is the beaming angle of the GWB signal is given by \cite{Damour:2001bk} 
\begin{equation}
\theta^{div}(f,z)=\left(\frac{4}{\sqrt{3}g_2(1+z)fl}\right)^{1/3}\simeq \left(g_2(1+z)fl\right)^{-1/3},
\label{thetadivprop}
\end{equation}
where the dimensionless parameter $g_2$ is defined as
\begin{equation}
g_2=\left(min\left(|\vec{a}''(u_c,v_c)|,|\vec{b}''(u_c,v_c)|\right)\right)^{-1}.
\label{g2}
\end{equation}
As with $g_1$, $g_2$ has typically been assumed to be of order unity \cite{Damour:2001bk, Abbott:2017mem}, and in \cite{Pazouli:2020qmj} we confirmed this is a good approximation by considering a class of odd harmonic loops which led to many cusp forming events, allowing us to obtain excellent statistics on the distribution of values for $g_1$ and $g_2$.
The restriction applied by the angle $\theta^{div}$ is that the GWB is only observed if the angle between the velocity of the cusp and the direction of the observer is less than $\theta^{div}$.

Note that although the cosmic string loop emits at discrete frequencies $f_m$, where $m$ is the harmonic mode, we consider a high frequency continuous approach to reach the GWB amplitude expression \cite{Damour:2001bk}. Then, the low frequency limit $m\simeq 1$, i.e. $f_m=f_{em}=(1+z)f\simeq T_1^{-1}$ (where $f_{em}$ is the frequency of the emitted GWB at the source), of the GWB amplitude would be 
\begin{equation}
\tilde{h}^{cusp}_{LF}(z)\simeq  Cg_1G\mu l \frac{1}{r(z)},
\label{hlowfreq}
\end{equation}
and therefore, the high frequency amplitude compared to the low frequency one is $\tilde{h}^{cusp}(z)\simeq \theta_{m}\tilde{h}^{cusp}_{LF}(z)$, where 
\begin{equation}
\theta_{m}\simeq \left((1+z)fl\right)^{-1/3}\simeq m^{-1/3}.
\label{thetam}
\end{equation}
Since the cosmic string loop emits at frequencies $f_m$ with $|m|\geq 1$, the condition $\theta_m\leq 1$ should hold for the continuous limit. This restriction is necessary to make sure that we do not take into account non-existing modes with $|m|\leq 1$.

\subsection{Set of values for $g_1$, $g_2$ and $c$}\label{g12parameters}
In this section we will estimate the values of $g_1$ and $g_2$ using our results for the second derivatives of the right- and left-movers at the cusp occuring on odd-harmonic cosmic strings, $|\vec{a}''(u_c,v_c)|$ and $|\vec{b}''(u_c,v_c)|$, which were calculated in terms of the harmonic order in \cite{Pazouli:2020qmj}. Combining them with the results we obtained for our toy model in section \ref{toymodelevol} we will obtain sets of values for $g_1$, $g_2$ and $c$, by assuming the harmonic order distribution of the parent loops in a unit spacetime volume.   

In \cite{Pazouli:2020qmj} we concluded that the values of $g_1$ and $g_2$ are of order unity for the odd-harmonic family of strings, with their average value increasing slowly as the harmonic order increases. This was calculated by considering a large number of loops of each harmonic order from $N=1$ up to $N=21$. In particular, we found that $0.38 < g_1 < 1$ and $0.34< g_2 <1$. Since the values of the second derivatives do not differ a lot with respect to the harmonic order, an approach to estimate $g_1$ and $g_2$ would be to calculate them by averaging over the complete set of values for $g_1$ and $g_2$ at each cusp event regardless of the harmonic order. The only assumption of this model is that the harmonic order distribution ranges from $N=1$ to $N=21$, without assuming the percentage of parent loops at each harmonic order. Averaging over a total of 278069 cusp events, we find that $g_1=0.489$ and $g_2=0.305$. In particular, we averaged over 30000 cusp events from each harmonic order, except the 21st harmonic order where we used 6699 events and the 23rd harmonic order where we used 1370. This reduction in the number of events in higher harmonic order loops is because of the increased computation time the higher harmonic order loops require to be analyzed. This result of $g_1$ suppresses the GWB amplitude \eqref{hcuspprop2} by around half, compared to the estimation of \cite{Damour:2001bk}, while the observation angle of the GWB from cusps, which is $\theta^{div}$, will increase by a factor of $1.5$ compared to \cite{Damour:2001bk}. For this estimation, we will use $c=1$, which is the value used in \cite{Damour:2001bk,Abbott:2017mem} for the average cusps per period. Thus, one of the sets of values that we will use for the estimation of the GWB amplitude is $(g_1=0.489, g_2=0.305,c=1)$, which we will call set 1. We will call set 0 the values $(g_1=1,g_2=1,c=1)$, which is the set of values assumed in both \cite{Damour:2001bk} and \cite{Abbott:2017mem}.

To estimate values of $c\neq 1$, we will use our assumptions of the harmonic order distribution of the parent loops in a spacetime volume $dV(z)$, which were presented in section \ref{Cuspsperperiodnetwork}. Using this distribution of harmonics, we will also calculate the corresponding average values of $g_1$ and $g_2$. Note that the distribution of the harmonics is assumed to include parent loops of harmonic orders from $N=1$ to $N=7$. For the uniform distribution case, we find that $(g_1=0.680,g_2=0.699,c=8.82)$, for Benford's law we find $(g_1=0.773,g_2=0.779,c=6.43)$ and for the geometric distribution, $(g_1=0.839,g_2=0.842,c=4.0)$. For ease of viewing, these model values are summarised in Table~\ref{tableg1g2csummary}.

\begin{table}
\centering
\begin{tabular}{ |c||c|c|c|c|c|  }
 \hline
 Model    & $g_1$ & $g_2$ & $c$\\
 \hline
 Set 0 (LIGO/VIRGO Ref\cite{Abbott:2017mem})& 1.0   & 1.0 & 1.0 \\
 Set 1 (Ref\cite{Pazouli:2020qmj}) & 0.489   & 0.305 & 1.0  \\
 Set 2 (Uniform) & 0.680 & 0.699 & 8.82  \\ 
 Set 3 (Benford) & 0.773 & 0.779 & 6.43  \\
 Set 4 (Geometric) & 0.839 & 0.842 & 4.0 \\
\hline
\end{tabular}
\caption{The values of $g_1,g_2$ and $c$ derived in Section~\ref{g12parameters} for the five models being considered when evaluating the GWB event rate from cosmic strings.}
\label{tableg1g2csummary}
\end{table}

\section{Rate of GWBs from a cosmic string network}\label{GWBcs}
In this section we will calculate the GWBs observed on Earth using the semi-analytic Model 1 from \cite{Abbott:2017mem}, which we have presented in section \ref{CosmicStringNet}. Our aim is to calculate the rate of GWBs emitted from cusps on cosmic strings arriving to Earth for this cosmic string network using the odd-harmonic string assumptions presented in \ref{g12parameters}, and to compare our results with those presented in \cite{Abbott:2017mem}.

\subsection{Calculation of the GWB event rate}\label{dRnumer}
We can write the length $l$ of a loop in terms of the amplitude $h$, the frequency $f$ and the redshift $z$
\begin{equation}
l(h,z,f)=\left(\frac{hf^{4/3}(1+z)^{1/3}\varphi_r(z)}{g_1G\mu H_0}\right)^{3/2}.
\label{Linhzf}
\end{equation}
The above is obtained by inverting equation \eqref{hcuspprop2}. We can also express $\theta^{div}$ in terms of $h$, $f$ and $z$. By combining equations \eqref{thetadivprop} and \eqref{Linhzf}, we find
\begin{equation}
\theta^{div}(h,f,z)=\left[g_2\left(\frac{f^2h(1+z)\varphi_r(z)}{g_1G\mu H_0}\right)^{3/2}\right]^{-1/3}.
\label{thetadivh}
\end{equation}
Finally, we define the number of cusps per unit space time volume and for GWBs of amplitudes between $h$ and $h+dh$ \cite{Abbott:2017mem} 
\begin{equation}
\nu(h,z,f)dh=\nu(l(h,z),z)\frac{dl}{dh}dh=\nu(l(h,z),z)\frac{3}{2h}ldh
\label{nudhdl}
\end{equation}
where we used the quantity $\nu(h,z,f)$ defined in equation \eqref{cuspsdv}.

We can define the rate of GWBs in the unit spacetime volume $dV(z)$ and in an interval of amplitudes from $h$ to $h+dh$, \cite{Abbott:2017mem} 
\begin{equation}
\frac{d^2R}{dV(z)dh}=\left(\frac{\theta^{div}(h,z,f)}{2}\right)^2(1+z)^{-1}\nu(h,z,f)\Theta(1-\theta^{div}(h,z,f)).
\label{dRdVdh}
\end{equation}
We use equation \eqref{nudhdl} to change the coordinates of $\nu$ from $h$ to $l$ in the above expression, and find that
\begin{equation}
\begin{aligned}
\frac{d^2R}{dV(z)dh}=&\left(\frac{\theta^{div}(h,z,f)}{2}\right)^2(1+z)^{-1})\frac{3}{2h}l(h,z,f)\nu(l,z,f)\\
&\Theta(1-\theta^{div}(h,z,f))\\
=&\left(\frac{\theta^{div}(h,z,f)}{2}\right)^2(1+z)^{-1})\frac{3}{2h}l(h,z,f)\\
&\frac{2}{l(h,f,z)}c\, n(h,z,f)\Theta(1-\theta^{div}(h,z,f)).
\label{}
\end{aligned}
\end{equation}
Using equation \eqref{dVzdef}, we find the derivative of $R$ in terms of the redshift 
\begin{equation}
\begin{aligned}
\frac{d^2R}{dzdh}=\left(\frac{\theta^{div}(h,z,f)}{2}\right)^2\frac{3c\varphi_V(z)}{H_0^{3}(1+z)ht(z)^4}\mathcal{F}(l,z,f)
\Theta(1-\theta^{div}(h,z,f)).
\label{}
\end{aligned}
\end{equation}
Finally, substituting $\theta^{div}$ from equation \eqref{thetadivh}, we reach the expression
\begin{equation}
\begin{aligned}
\frac{d^2R}{dzdh}(h,f,z)=&\frac{3}{4}\frac{g_1}{g_2^{2/3}}\frac{G\mu c\varphi_V(z)}{H_0^2 f^2 (1+z)\varphi_r(z) t(z)^4}\frac{1}{h^2}\mathcal{F}(l,z)\\
&\Theta(1-\theta^{div}(h,f,z)).
\label{dRdzdh}
\end{aligned}
\end{equation}
The above expression is true for any cosmological era, i.e. for all redshifts, and for any cosmic string model.

We will now proceed to calculate the rate of GWBs separating the matter era calculation from the radiation
era calculation (i.e. we will no longer use interpolating functions between the two eras). During the radiation era, $z>z_{eq}=3366$, we substitute equation \eqref{calFraddef} into the GWB rate \eqref{dRdzdh} to find that the rate of GWBs during the radiation era is 
\begin{equation}
\begin{aligned}
\frac{d^2R_{rad}}{dzdh}(h,f,z)=&3\frac{g_1}{g_2^{2/3}}\frac{G\mu \pi c}{ f^2 }\frac{H_0^2\varphi_r(z)}{(1+z)^5\varphi_t(z)^4\mathcal{H}(z)}\frac{1}{h^2}\\&\frac{C_{rad}}{\left[\frac{H_0}{\varphi_t(z)}\left(\frac{hf^{4/3}(1+z)^{1/3}\varphi_r(z)}{g_1G\mu H_0}\right)^{3/2}+\Gamma G\mu\right]^{5/2}}\\
&\Theta\left(1-\theta^{div}(h,f,z)\right) \Theta\left(\alpha-\gamma(h,f,z)\right),
\label{dRraddzdh}
\end{aligned}
\end{equation}
where we used equation \eqref{phiVdef} to substitute $\varphi_V(z)$ with $\varphi_r(z)$ and equation \eqref{phitdef} to substitute $t(z)$ with $\varphi_t(z)$. Also, recall that $\alpha \sim 0.1$ is a constant determining the size of the loops formed from the long string network, and $\mathcal{H}(z)$ is the Hubble constant at redshift $z$ normalized by $H_0$, defined in equation \eqref{Hconstdef}. 

During the matter era, $z<z_{eq}=3366$, we substitute equation \eqref{calFmatdef} into the GWB rate \eqref{dRdzdh} to find that the rate of GWBs during the matter era is
\begin{equation}
\begin{aligned}
\frac{d^2R_{mat}}{dzdh}(h,f,z)=\frac{d^{2}R_{mat}^{(1)}}{dzdh}(h,f,z)+\frac{d^{2}R_{mat}^{(2)}}{dzdh}(h,f,z),
\label{dRmatdzdh}
\end{aligned}
\end{equation}
where the first term on the right-hand-side of the equation corresponds to the GWBs originating from loops that formed in the radiation era and survive into the matter era, and the second term on the right-hand-side of the equation corresponds to the GWBs originating from matter era loops. They are given by
\begin{equation}
\begin{aligned}
\frac{d^{2}R_{mat}^{(1)}}{dzdh}(h,f,z)=&3\frac{g_1}{g_2^{2/3}}\frac{G\mu \pi c}{f^2 }\frac{H_0^2\varphi_r(z)}{(1+z)^5\varphi_t(z)^4\mathcal{H}(z)}\frac{1}{h^2}\left(\frac{\varphi_t(z_{eq})}{\varphi_t(z)}\right)^{1/2}\\
& \frac{C_{mat}}{\left[\frac{H_0}{\varphi_t(z)}\left(\frac{hf^{4/3}(1+z)^{1/3}\varphi_r(z)}{g_1G\mu H_0}\right)^{3/2}+\Gamma G\mu\right]^{2}}\\
& \Theta\left(1-\theta^{div}(h,f,z)\right) \Theta\left(-\gamma(h,f,z)+\beta(t)\right),
\label{dRmat1dzdh}
\end{aligned}
\end{equation}
and
\begin{equation}
\begin{aligned}
\frac{d^2R_{mat}^{(2)}}{dzdh}(h,f,z)=&3\frac{g_1}{g_2^{2/3}}\frac{G\mu \pi c}{f^2 }\frac{H_0^2\varphi_r(z)}{(1+z)^5\varphi_t(z)^4\mathcal{H}(z)}\frac{1}{h^2}\\
& \frac{C_{mat}}{\left[\frac{H_0}{\varphi_t(z)}\left(\frac{hf^{4/3}(1+z)^{1/3}\varphi_r(z)}{g_1G\mu H_0}\right)^{3/2}+\Gamma G\mu\right]^{5/2}}\\
& \Theta\left(1-\theta^{div}(h,f,z)\right) \Theta\left(\gamma(h,f,z)-\beta(t)\right)\\
& \Theta\left(\alpha-\gamma\right),
\label{dRmat2dzdh}
\end{aligned}
\end{equation}
respectively. Note that 
\begin{equation}
\begin{aligned}
\gamma(z)=\frac{l(z)}{t(z)}=\frac{H_0}{\varphi_t(z)}\left[\frac{hf^{4/3}(1+z)^{1/3}\varphi_r(z)}{g_1G\mu H_0}\right]^{3/2}
\label{gammaz}
\end{aligned}
\end{equation} 
and 
\begin{equation}
\begin{aligned}
\beta(z)=\alpha \frac{\varphi_t(z_{eq})}{H_0}-\frac{\Gamma G \mu}{H_0} \left(\varphi_t(z)-\varphi_t(z_{eq})\right),
\label{betaz}
\end{aligned}
\end{equation} 
as can be seen from equations \eqref{gammarelsize} and \eqref{betalength}.
The rate of GWBs is obtained by integrating over the redshift and the amplitude of the GWBs, and it is given by the integral
\begin{equation}
\begin{aligned}
R(h,z)=\int_0^{z_{max}}\int_{h_{min}(z)}^{h_{max}(z)}\frac{d^2 R(h,f,z)}{dzdh} dzdh
\label{Roverall}
\end{aligned}
\end{equation} 

The range of integration for $h$ is limited by the conditions we have imposed. The beaming angle of the cusps satisfies $\theta^{div}<1$, providing the lower limit $h_{min}(z)$
\begin{equation}
\begin{aligned}
h(z)>h_{min}(z)=\frac{g_1 G\mu H_0}{g_2^{2/3}(1+z)f^2\varphi_r(z)},
\label{hmin}
\end{aligned}
\end{equation} 
and holds for any era. During the radiation era, the upper limit $h_{max}(z)$ comes from considering the evolution of $h(z)$, when $\alpha>\gamma$ holds leading to 
\begin{equation}
\begin{aligned}
h(z)<h_{max}(z)=\frac{g_1 G\mu H_0\alpha^{2/3}t(z)^{2/3}}{f^{4/3}(1+z)^{1/3}\varphi_r(z)}.
\label{hmaxrad}
\end{aligned}
\end{equation} 
Turning our attention to the upper limit of the integral over the redshift, $z_{max}$, it can be obtained by combining equations \eqref{hmin} and \eqref{hmaxrad}, where we find
\begin{equation}
\begin{aligned}
\varphi_t(z)(1+z)\geq \frac{H_0}{\alpha g_2 f}.
\label{}
\end{aligned}
\end{equation} 
As we are working in the regime where $z\gg 1$, we can use the large redshift expression for $\varphi_t(z)$, given in equation \eqref{phitgg1} which simplifies things considerably. Since the value of $z_{max}$ varies with $g_2$, which we determine numerically for the various models we considered in section~\ref{g12parameters}, we have presented its value with respect to $g_2$ in Table \ref{tabzmaxrad}.
\begin{table}
\centering
\begin{tabular}{ |c||c|c|c|c|c|  }
 \hline
 Value    & $z_{max}$ \\
 of $g_2$ &           \\
 \hline
 Set 0, $g_2=1.000$   & $3.81\times 10^{20}$  \\
 Set 1, $g_2=0.305$   & $1.16\times 10^{20}$  \\
 Set 2, $g_2=0.699$   & $2.66\times 10^{20}$  \\ 
 Set 3, $g_2=0.779$   & $2.97\times 10^{20}$  \\
 Set 4, $g_2=0.842$   & $3.21\times 10^{20}$  \\

 \hline
\end{tabular}
\caption{The values of the limit of integration $z_{max}$ in the radiation era for each value that $g_2$ takes in the set 0 to set 4.}
\label{tabzmaxrad}
\end{table}

Turning our attention now to the matter era, we need to consider both the limits of integration for the loops surviving from the radiation era, as well as those for the loops formed in the matter era. For the former the inequality 
\begin{equation}
\begin{aligned}
-\gamma(z)+\beta(z)\geq 0
\label{gammabetaconstraint}
\end{aligned}
\end{equation}
holds, restricting the loop length within the limits of loops formed in the matter era. Using equations \eqref{gammaz}-\eqref{betaz}, and substituting $l(h,z,f)$ from equation \eqref{Linhzf},  \eqref{gammabetaconstraint} becomes
\begin{equation}
\begin{aligned}
h\leq h_{max}=\frac{g_1 G\mu H_0^{1/3}}{f^{4/3}(1+z)^{1/3}\varphi_r(z)}\left(\varphi_t(z)\beta(z)\right)^{2/3}.
\label{hmaxmat1}
\end{aligned}
\end{equation}
Recalling $\beta(z) >0$ as it is the length of a loop at time $t(z)$, which was formed at $t(z_{eq})$, and requiring that $h_{min}$ (given by equation \eqref{hmin}) is less than $h_{max}$, we find that 
\begin{equation}
\begin{aligned}
0 \leq -\frac{H_0}{g_2 f (1+z)}+ (\alpha+\Gamma G \mu)\varphi_t(z_{eq})-\Gamma G \mu \varphi_t(z).
\label{zminmat1}
\end{aligned}
\end{equation}
The inequality \eqref{zminmat1} provides us with a value $z_{min}$, where $z_{min}$ is the redshift where all the loops that formed in  the radiation era but survived into the matter era vanished. Solving \eqref{zminmat1} numerically, we find that the $g_2$ term is small compared to the remaining terms, implying little sensitivity in $z_{min}$ to $g_2$. We find that $z_{min}=0.288$ in the matter era. Finally, for the loops formed in the matter era, the relevant inequalities become $1-\theta_m\geq 0$, $\gamma(z)-\beta(z)\geq 0$ and $\alpha-\gamma(z)\geq 0$. Combining these we obtain the constraints on $h_{min}$ namely
\begin{equation}
\begin{aligned}
h\geq h_{min}^{(1)}=h_{min},
\label{}
\end{aligned}
\end{equation}

\begin{equation}
\begin{aligned}
h\geq h_{min}^{(2)}=\frac{g_1 G\mu H_0^{1/3} \varphi_t(z)^{2/3}\beta(z)^{2/3}}{f^{4/3}(1+z)^{1/3}\varphi_r(z)},
\label{}
\end{aligned}
\end{equation}

\begin{equation}
\begin{aligned}
h \leq h_{max}=\frac{g_1 G\mu H_0^{1/3}\alpha^{2/3}\varphi_t(z)^{2/3}}{f^{4/3}(1+z)^{1/3}\varphi_r(z)},
\label{}
\end{aligned}
\end{equation}
respectively. We find that $h_{min}^{(1)}\geq h_{min}^{(2)}$ for all z in the matter era, by plotting both functions. Therefore, the crucial constraint on $h_{min}$ is  $h\geq h_{min}^{(1)}=h_{min}$.

\section{Results}\label{results}
We can now determine the rate of GWBs from cusps on cosmic strings in redshift intervals defined by
\begin{equation}
\begin{aligned}
R(z)=\int_z^{z+\Delta z}\int_{h_{min}(z')}^{h_{max}(z')}\frac{d^2 R}{dz'dh}dhdz'
\label{Rzplotdef}
\end{aligned}
\end{equation}
where from now on, we choose the frequency to be $f=100Hz$, similarly to that used in \cite{Abbott:2017mem}. The range of redshifts appearing in the figures will be divided into 1000 intervals ranging from $z_1=10^{-12}$ to $z_2=10^{32}$, 
in equation \eqref{Rzplotdef}. $\Delta z$ is the interval width at redshift z, and the integration has a lower limit of 
\begin{equation}
\begin{aligned}
z=z_1\left(\frac{z_2}{z_1}\right)^{\frac{b}{n}}
\label{}
\end{aligned}
\end{equation}
and an upper limit of
\begin{equation}
\begin{aligned}
z+\Delta z=z_1\left(\frac{z_2}{z_1}\right)^{\frac{b+1}{n}},
\label{}
\end{aligned}
\end{equation}
where the counter $b$ takes integer values in the interval $0\leq b\leq n$ and $n=1000$. Note that the functional form of $d^2R/dzdh$, as well as the limits of integration $h_{min}(z)$ and $h_{max}(z)$, change with the cosmological era, matter or radiation. Therefore, we will deal with the rate of GWBs from the loops formed in the radiation era, the loops surviving into the matter era and the loops formed in the matter era separately.

For the loops formed in the radiation era, we will integrate the function in equation \eqref{dRraddzdh}, with limits of integration $h_{min}(z)$ and $h_{max}(z)$ given by equations \eqref{hmin} and \eqref{hmaxrad} respectively, and with the redshift ranging from $z=z_{eq}=3366$ to $z=z_{max}$, which depends on the value of $g_2$. This corresponds to integer values of the counter $b$ from 352 to roughly 476 (depending on the value of $g_2$). The integral of $d^2R/dzdh$ over $h$ is calculated analytically. After integrating it we obtain the function 
\begin{equation}
\begin{aligned}
\frac{dR_{rad}}{dz}(z,h_{max}(z))-\frac{dR_{rad}}{dz}(z,h_{min}(z)),
\label{GWBratediff}
\end{aligned}
\end{equation} 
which is a function of redshift only. It is not possible to integrate this function analytically over $z$, so this is done numerically over the redshifts in the intervals given by \eqref{Rzplotdef}. Since we calculate this for redshifts $z\gg 1$, we use the asymptotic expressions for $\varphi_t(z)$ and $\varphi_r(z)$, given by equations \eqref{phitgg1} and \eqref{phirgg12}, respectively. For $3366\leq z\leq 10^9$ (i.e. $b\in [352,476]$), we set $\mathcal{G}(z)=1$ (see equation \eqref{Gcaldef}), and we numerically integrate the above function over $z$, thus obtaining the plot of the rate of GWBs \eqref{Rzplotdef} for this range of redshifts. We apply the same method for $10^9<z\leq 2\times 10^{12}$ (i.e. $477\leq b\leq 551$), setting $\mathcal{G}=0.83$. Finally, we move to the region of redshifts $2\times 10^{12}<z\leq z_{max}$, which corresponds to the integral bins $552\leq b\leq 739$. 
This is a challenging regime to work in, as the two leading terms in equation  \eqref{GWBratediff} are within $10^{-12}$ of each other, leading to numerical errors dominating the solution, which only increase with increasing $z$. This issue appears for $z>10^{16}$ in particular. To resolve it we increase the maximum machine precision in Mathematica, ensuring it calculates all the digits that are significant for the function \eqref{GWBratediff}. Then, we apply a fifth order Taylor series around a number of redshift points, between $z=2\times 10^{18}$ and $2 \times 10^{20}$ allowing us to resolve the issue of numerical accuracy. When integrating over $z$, we use the expressions obtained from the Taylor series to obtain the final result, which is the rate of the GWBs given in equation \eqref{Rzplotdef}. In this way, we can finally obtain the plot of $R(z)$ in the radiation era.

During the matter era, we have two types of loops, the ones that formed in the radiation era and survived into the matter era, which are given by equation \eqref{dRmat1dzdh}, and the ones formed in the matter era, which are given by equation \eqref{dRmat2dzdh}. We plot these in separate figures over the redshifts $[0.288,3366]$, which correspond to the $b$ values in the interval $[261,352]$. Note that during the matter era we cannot use the asymptotic expressions for $\varphi_t(z)$ and $\varphi_r(z)$, apart from $z\ll 1$. Therefore the equations are solved numerically, applying the following procedure. First, we create a list of the redshift values for each interval from $z$ to $z+\Delta z$ in the matter era, which consists of the points
\begin{equation}
\begin{aligned}
z_i=z+i \frac{\Delta z}{n}
\label{}
\end{aligned}
\end{equation} 
where $i$ takes on integer values such that $0\leq i\leq n$. We can easily integrate equation \eqref{GWBratediff} for each value $z=z_i$ using for example Simpson's rule, leading to the desired result  \eqref{Rzplotdef}.

\begin{figure}%
\centering
\includegraphics[scale=0.65]{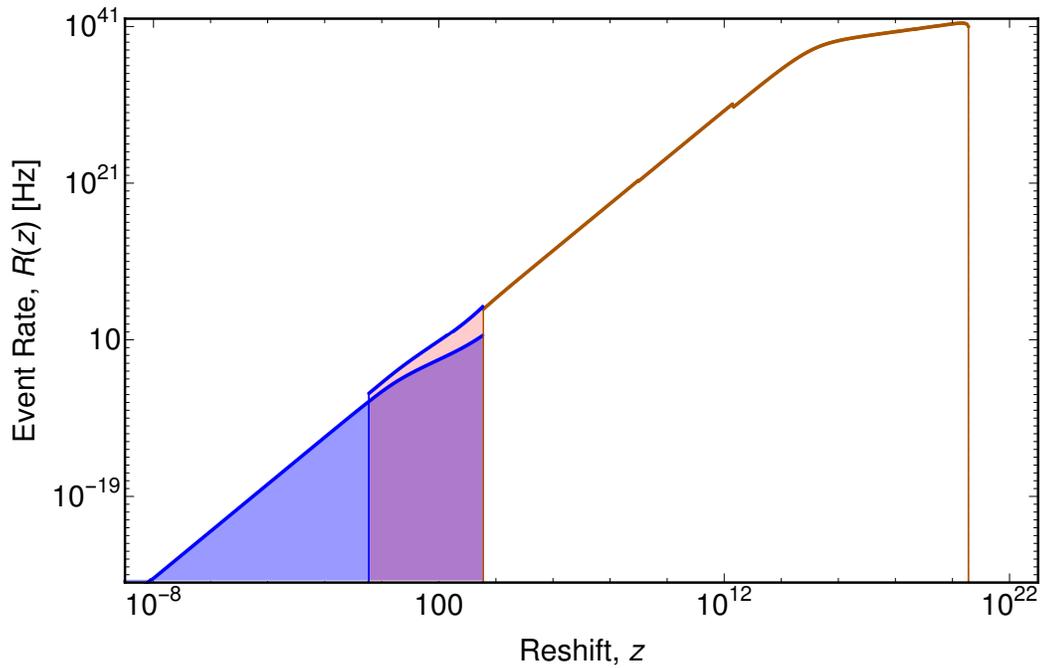} %
\caption{The plot of the GWB event rate, $R(z)$, versus redshift, $z$, for $G\mu=10^{-8}$, $g_1=g_2=1$, $c=1$ and $f=100 Hz$, for the ``one-scale" cosmic string network model.}%
\label{Ligoplot}%
\end{figure}

Figure \ref{Ligoplot} shows the event rate of GWBs, given in equation \eqref{Rzplotdef}, versus redshift. The orange line corresponds to the event rate during the radiation era ($z_{eq}=3366 < z < z_{max}$). The matter era contribution, ($10^{-8}< z< z_{eq}$), is shown by both the light red shaded plot, which corresponds to the event rate arising from radiation era formed loops, and the blue shaded plot, which corresponds to the event rate from matter era formed loops. This figure has been reproduced using the values of $g_1,g_2$ and $c$ or Set 0, which are the same assumptions as used in \cite{Abbott:2017mem}, and it corresponds to the upper left-hand-side plot of their Figure 7. Note that the blue shaded region that overlaps with the light red region differs in our plot compared to theirs. However, this does not affect the event rate as it is not the dominant contribution of GWBs, since the event rate from radiation loops surviving into matter is stronger by at least one order of magnitude.  

\begin{figure}%
\centering
\includegraphics[scale=0.65]{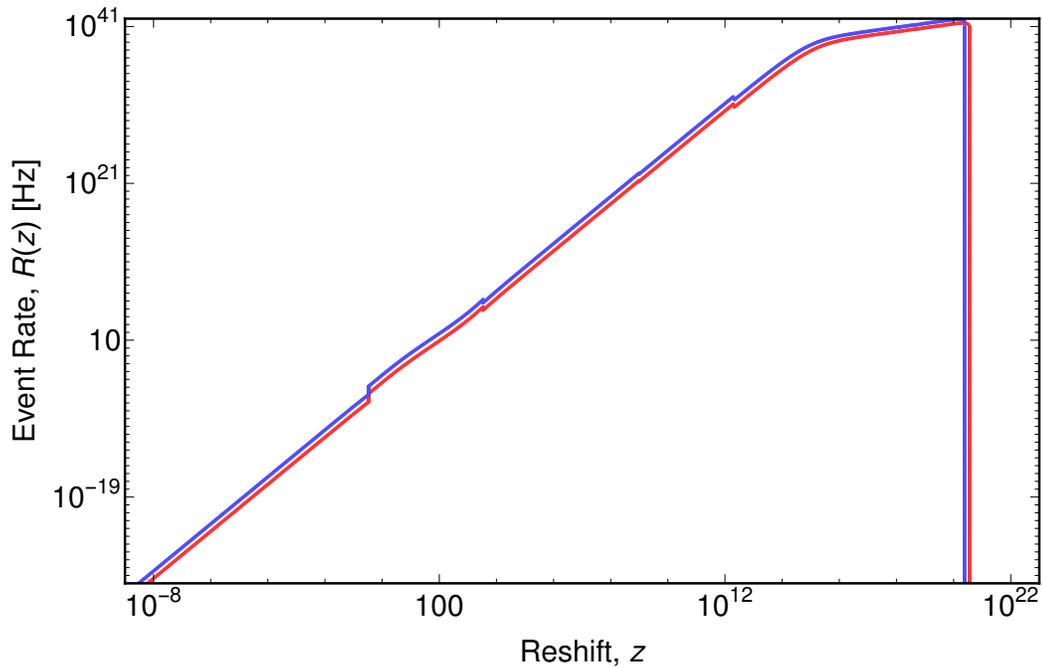} %
\caption{The plot of the GWB event rate, $R(z)$, versus redshift, $z$, for $G\mu=10^{-8}$ and $f=100 Hz$. The red plot corresponds to the set 0 values $g_1=g_2=1$, $c=1$ and the blue plot corresponds to the set 2 values $(g_1=0.680,g_2=0.699,c=8.82)$ and shows an increase of roughly one order of magnitude over the set 0 results.}%
\label{Plotset02}%
\end{figure}

Finally we turn our attention to how the results of the GWBs are modified in our new models. The impact is shown in Figure \ref{Plotset02}. In particular we plot the event rate for different values of $g_1$, $g_2$ and the cusp number, $c$, based on the results of our Toy model of section \ref{toymodelevol} and \ref{g12parameters}, and summarised in Table~\ref{tableg1g2csummary}. Out of the three sets of values $(g_1,g_2,c)$ presented in \ref{g12parameters}, it is set 2 that has the greatest impact to the event rate compared to set 0 (the values used in \cite{Abbott:2017mem}). In particular, we can see in Figure \ref{Plotset02} that the change is an increase of one order of magnitude. Note that during the era where matter formed loops and radiation formed loops that survive into the matter era coexist, we have plotted the integral of the quantity on the left-hand-side of equation \eqref{dRmatdzdh}, unlike in figure \eqref{Ligoplot} where we have plotted the quantities on the right-hand-side of equation \eqref{dRmatdzdh} separately. The results that are provided using the set 1 values are effectively the same as the results of \cite{Abbott:2017mem}, which use the set 0 values. The results using the set 3 and set 4 values have a difference of around half an order of magnitude compared to the set 0 results.

\section{Conclusions}\label{concl}
The impact of the LIGO observations on the status of cosmic strings has been profound, with very tight constraints emerging on the string tension as seen for example in \cite{LIGOScientific:2021nrg}, where they found that $G\mu < 9.6\times 10^{-9}$ for their Model 1, which is the model that we use to compare our results with. When assessing the constraints, it is important to bear in mind that there is significant modelling involved, in this case in the specific properties of the cosmic string networks. This is the case when considering the emission of GWs from cusps forming on a network of cosmic strings. Two of the key parameters are known as $g_1$ and $g_2$ and are defined in terms of the second derivatives of the left and right movers on the string (see equations \eqref{g1def} and \eqref{g2}). They typically are taken to have values of order unity, and in an earlier paper, we analysed whether this was a good assumption by considering cusp production on a class of high harmonic cosmic strings \cite{Pazouli:2020qmj}, coming to the conclusion after analysing the properties of tens of thousands of cusp events that indeed the quantities $g_1$ and $g_2$ were consistent with unity. The third key parameter which had to be assumed in \cite{Abbott:2017mem,LIGOScientific:2021nrg} was $c$, the typical number of cusps to be found on a stable non-self intersecting loop. This was taken to be unity, but also values closer to 0.1 were also considered (see also \cite{Damour:2000wa,Damour:2001bk}). In \cite{Pazouli:2020qmj} we were unable to determine this number because our loops tended to self intersect and we were unable to follow their full evolution. An independent determination of this parameter, hence of the GWB signal from cusp events has been the motivation behind this paper. We have developed an algorithm with which we can estimate $c$. Our approach has not involved solving the dynamical equations of motion for a network, and trying to determine the resulting distribution of cusps. This is notoriously difficult to do in a way that allows one to keep control over all the relevant length scales and times involved. Rather, we have followed the idea first proposed in \cite{PhysRevD.36.987} and \cite{PhysRevD.33.872}, in which the authors considered a binary tree approach, where an initial loop of a given harmonic order would self-intersect according to certain probabilities which related to the size of the loop being produced and the length of the original loop etc... This probabilistic approach allowed them to estimate the type of loop configurations one could expect at the end of an evolution. We have modified this approach in this paper, proposing a new binary tree algorithm (see Section \ref{toymodelevol}) which leads to predictions for the number of stable non-self intersecting loops emerging from a given parent loop of odd harmonic number $N$. By slightly modifying the ansatz to consider only the smaller harmonic numbers ($N\leq 7$) we developed and analysed a series of toy models in which the cusp distribution on loops of non-self intersecting string can be determined. Of course our answers depend to some degree on the assumptions that we make, by far the most significant being that a given loop, if it intersected, did so as to produce two equal sized loops, with harmonic number reduced by 2 on each of the loops, but what is noticeable is the robust nature of the results. For example, we find that typically a network can effectively have of order 4-8 or so cusps per period, which is just under an order of magnitude larger than is assumed. Moreover, coupled with our previous results for the parameters $g_1$ and $g_2$, these results for $c$ have allowed us to compare our results for the GW bursts with those of \cite{Damour:2000wa,Damour:2001bk}. The key result is plotted in Figure \ref{Plotset02} where we see that the impact of the slightly larger cusp number is to increase the event rate of GWBs by just under an order of magnitude. Although this sounds a lot, in reality, given the range of values involved for the event rates it means that the usual assumption of $c\sim 1$ is probably a good working assumption. In many ways, this is a reassuring result. The vanilla model assumed to date and that has led to the published constraints seems robust. There are of course a number of places where this analysis could be improved upon. Allowing loops to chop off the network with arbitary sizes compared to the parent loop would be a good start, if somewhat challenging, and allowing for the higher harmonic modes in a more systematic way would be worth investigating, even though as we have argued, we do not believe they are likely to survive as such high harmonic loops self-intersect rapidly. 

\acknowledgements
AA and EJC acknowledge support from STFC grant ST/T000732/1. DP acknowledges support from the University of Nottingham Vice Chancellor's Scholarship.

\section{Appendix}
\subsection{Examples of tree evolution for low harmonics}\label{lowharmtree}
In this section we will provide an example of the toy model calculations presented in \ref{calculationanyharm} for cosmic string loops of harmonic order $N=1$, $N=3$ and $N=5$.

The simplest possible case for the binary tree is the case of the $N=1$ odd-harmonic string, which does not self-intersect and hence does not produce any daughter loops. For this case we know that the stable loops produced will always correspond to a single loop with average cusps per period $c_1=2$ (as is found in \cite{Pazouli:2020qmj}). 
\subsubsection{The $N=3$ parent loop case}
For the case of the $N=3$ odd-harmonic string, we have two possible states of evolution, hence two binary trees. The first tree is where the parent loop does not self-intersect, and the second is where the parent loop self-intersects, producing two $N=1$ daughter loops. 

To calculate the number of stable loops, we note that the binary tree in the case of no self-intersection has one leaf, i.e. one stable loop, and its probability of occurring is $\left[1-P(3,1)\right]=0.4$. The binary tree in the case of self-intersection has two leaves and its probability of occurring is $P(3,1)=0.6$. Averaging over the two possible cases we find that the average number of stable loops for $N=3$ is 
\begin{equation}
n_{s}^{(3)}=2P(3,1)+\left[1-P(3,1)\right]=1.6.
\label{N3stable}
\end{equation}

The lifetime of a loop is $\tau_l=l_i/\Gamma G\mu$. The period of the $N=3$ loop is $T_{3}^{(3)}=l/2=\alpha t_i/2$ . Note that we do not specify the value of $\alpha$, i.e. whether the loops are small or large, because it cancels. If it splits into two equal sized loops, each will have period $T_{1}^{(3)}=(l/2)/2=T_{3}^{(3)}/2$ (see equation \eqref{periodkl}). The two daughter loops produced will be of harmonic order $N=1$ and they will not split further. This means that they will live for a time $\tau = l_i/4\gamma G\mu$, since they lose energy with rate $\Gamma G\mu$ (see equation \eqref{lifetimekl}). Therefore, the average number of cusps emitted from the system per period of the initial loop defined in equations \eqref{gkl}-\eqref{cfinalk} consists of two terms, one with weight $P(3,3)=0.4$, which corresponds to the loop that does not self-intersect and one with weight $P(3,1)-0.6$, which corresponds to the self-intersecting loop. In case (a), the total number of cusps emitted from the system on average is
\begin{equation}
c_3\frac{\frac{\alpha t_i}{\Gamma G\mu}}{\frac{\alpha t_i}{2}}=c_3\frac{2}{\Gamma G\mu},
\label{N3t1a}
\end{equation}  
while the total number of periods is 
\begin{equation}
\frac{\frac{\alpha t_i}{\Gamma G\mu}}{\frac{\alpha t_i}{2}}=
\frac{2}{\Gamma G\mu},
\label{N3t1b}
\end{equation}
as we calculated in the previous section. Therefore, in the case that the loop does not split it produces $c_3$ cusps per period, as expected. In case (b), the total number of cusps formed on the system is
\begin{equation}
c_3\frac{\frac{\alpha t_i}{4}}{\frac{\alpha t_i}{2}}+2c_1\frac{\frac{\alpha t_i}{2\Gamma G\mu}}{\frac{\alpha t_i}{4}}=\frac{1}{2}c_3+\frac{4}{\Gamma G\mu}c_1.
\label{N3t2a}
\end{equation}
The first term on the right-hand side of \eqref{N3t2a}, corresponds to the average number of cusp events that occurred from the parent $N=3$ loop, in the half period interval before it splits, and the second term corresponds to the average number of cusp events that occurred from the two $N=1$ loops. Note that $c_3/2\ll 4c_1/\Gamma G\mu$ since $\Gamma G\mu$ is of the order $10^{-6}$ or smaller.
We can now calculate the average number of cusps emitted from the system of the $N=3$ loop per period of the initial loop, which is 
\begin{equation}
c^{(3)}=P(3,3)c_3+P(3,1)\left(\frac{1}{2}c_3+\frac{4}{\Gamma G\mu}c_1\right)\frac{\Gamma G\mu}{2}\simeq 4.8.
\label{N3cusppperiod}
\end{equation}
In the above, we have used the values $c_1=2$ and $c_3=5.96$ from Table 1 of \cite{Pazouli:2020qmj}.

\subsubsection{The $N=5$ parent loop case}

\begin{figure}%
    \centering
    \subfloat[The case where the parent loop does not chop into daughter loops.]{{\includegraphics[scale=0.2]{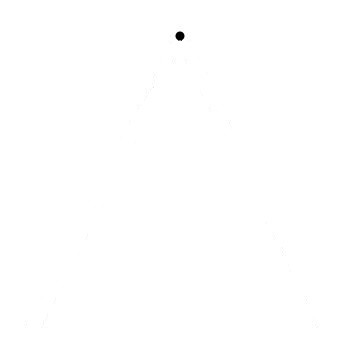} }}%
    \qquad
    \subfloat[The case where the parent loop chops into two third order harmonic loops.]{{\includegraphics[scale=0.2]{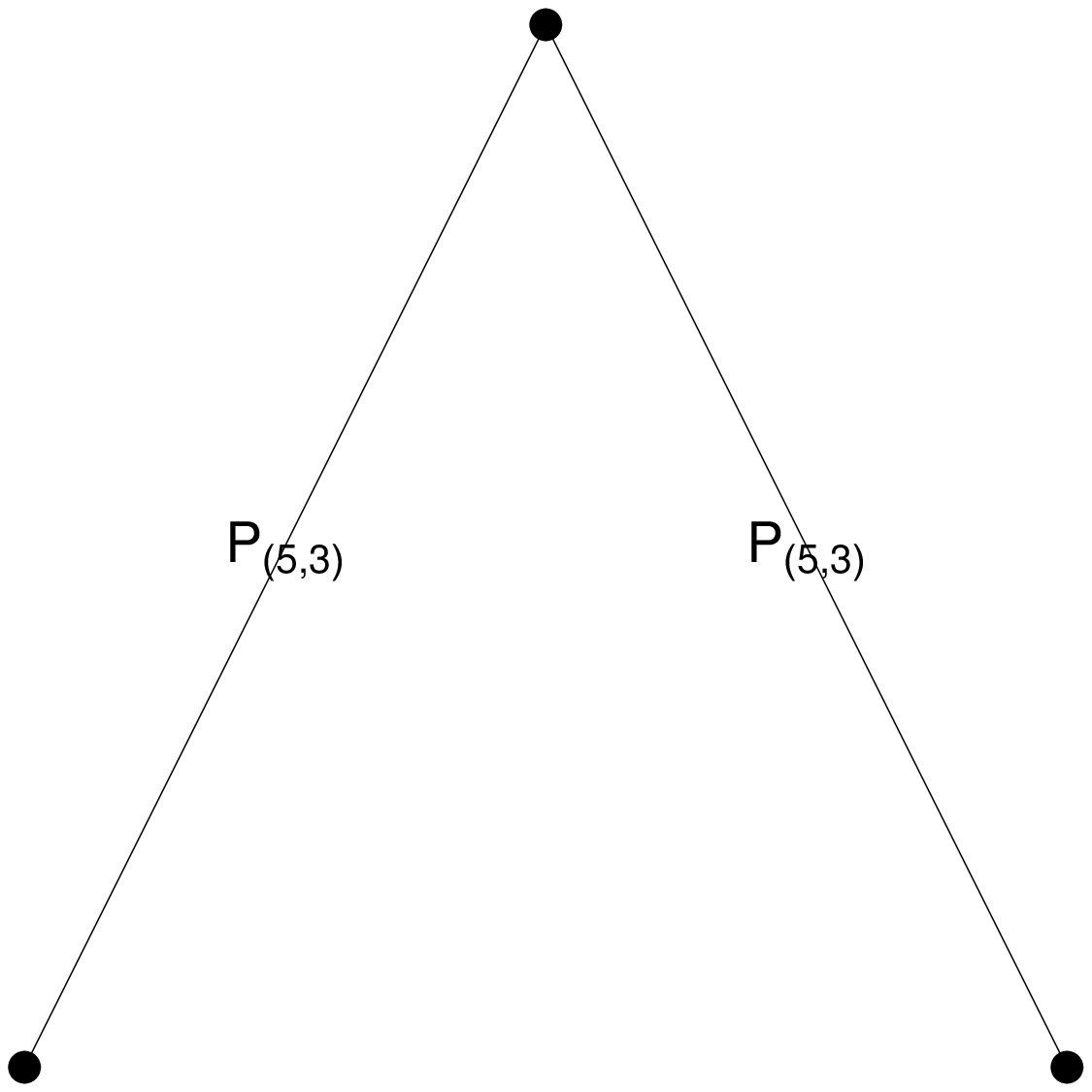} }}%
    \qquad
      \subfloat[The case where the parent loop chops into one third order and two first order harmonic loops.]{{\includegraphics[scale=0.2]{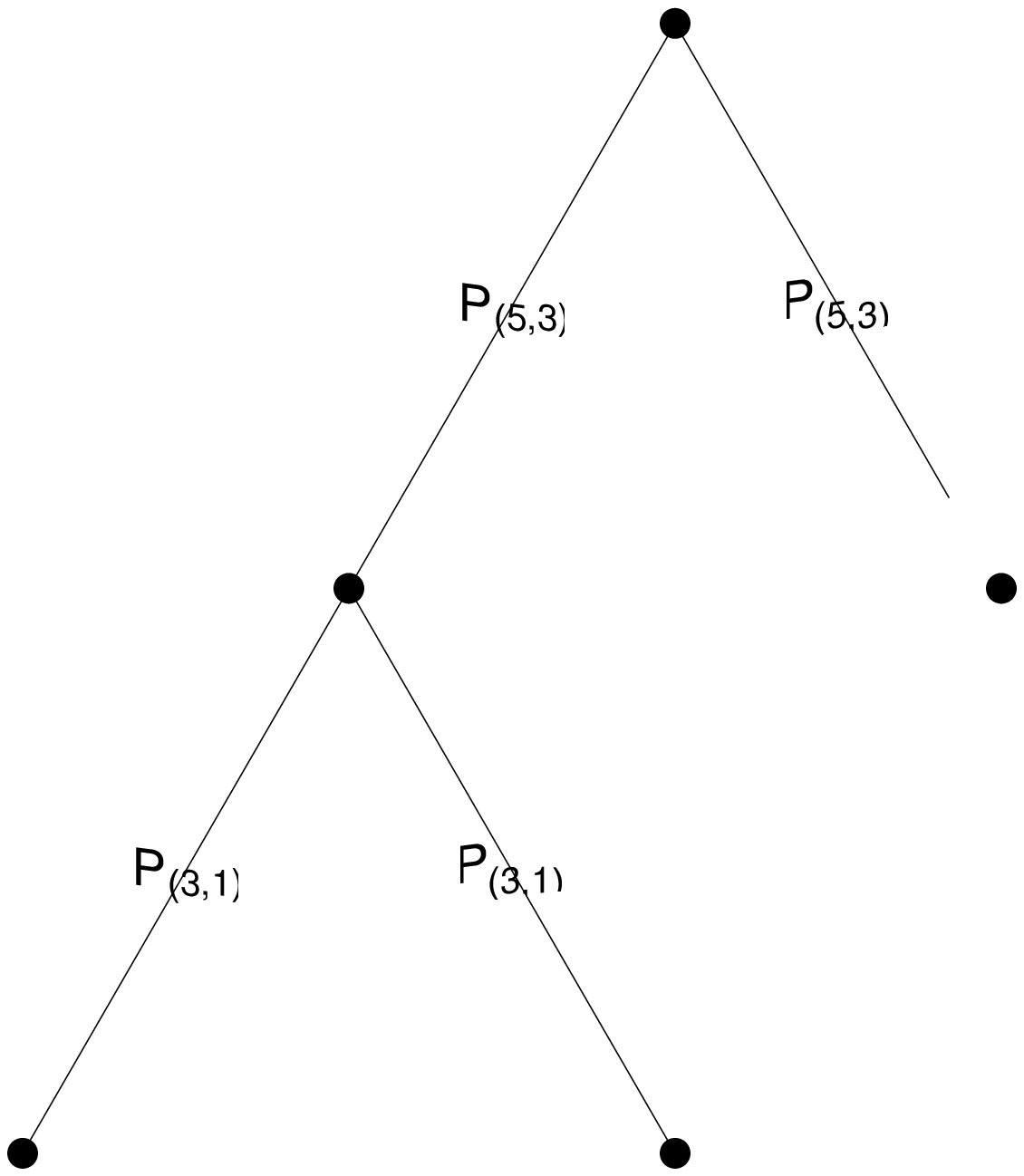} }}%
      \qquad
      \subfloat[The case which is the symmetric of the previous one.]{{\includegraphics[scale=0.2]{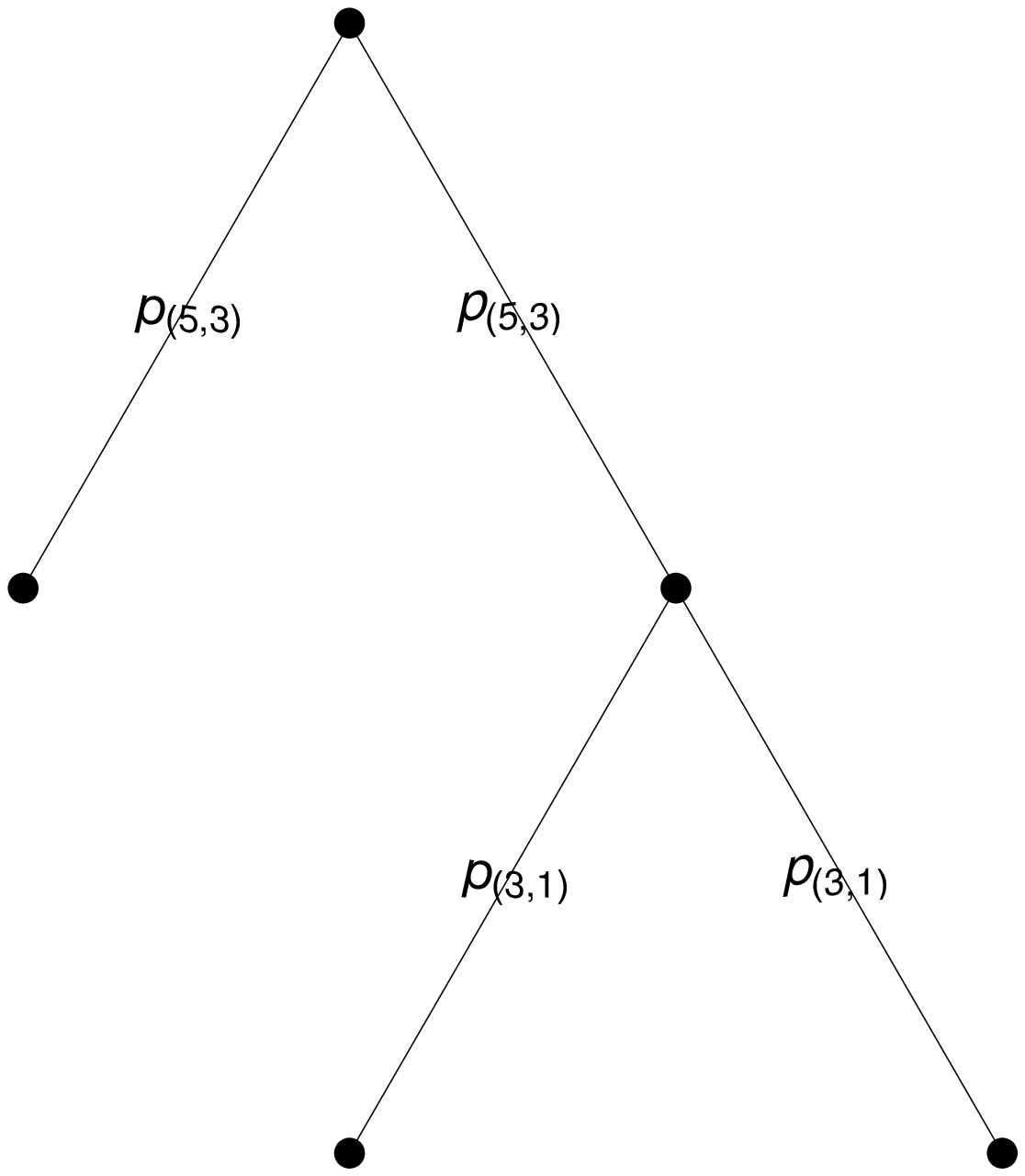} }}%
           \qquad
      \subfloat[The case where the parent loop chops into four first order harmonic loops.]{{\includegraphics[width=60pt, height=65pt]{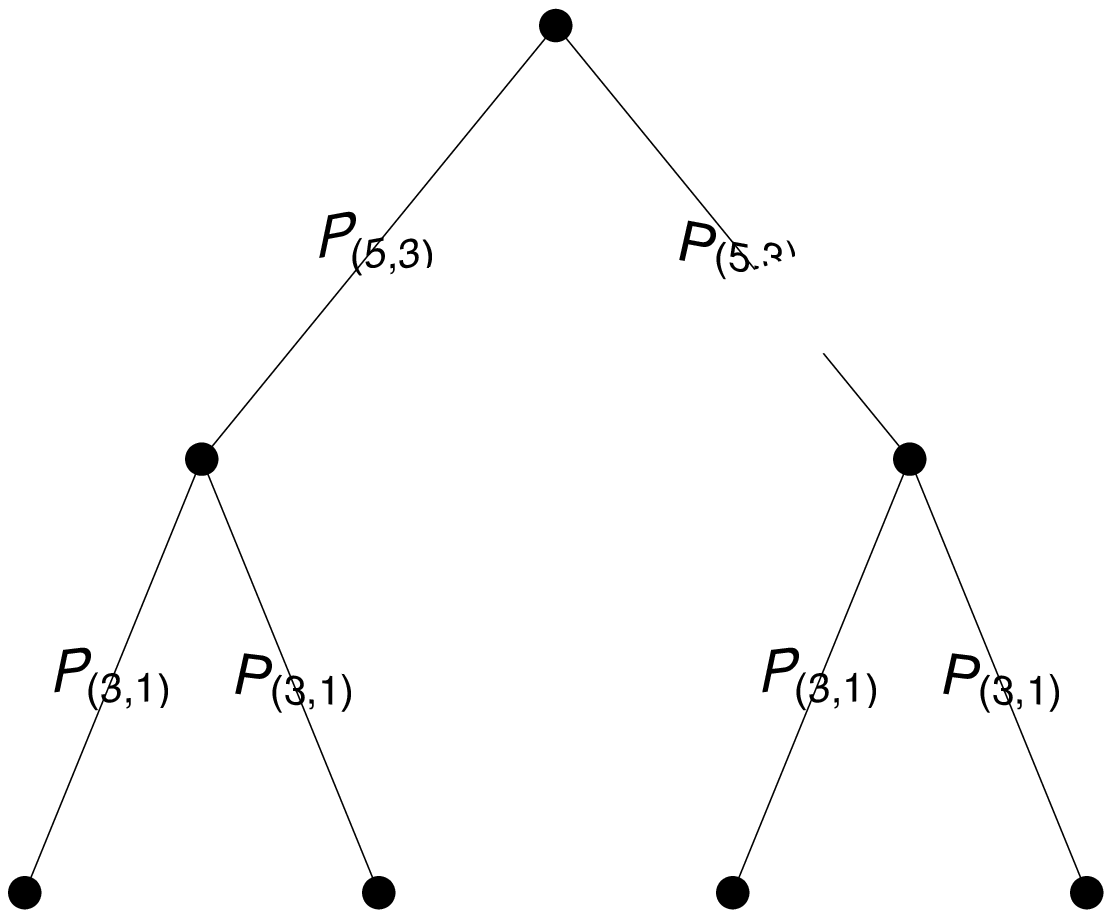} }}%
           
    \caption{All the possible cases of evolution for a fifth order harmonic string.}%
    \label{5harmonicevolution}%
\end{figure}

In the case of the fifth order odd-harmonic loop we have 5 configurations, as we can see in Figure \ref{5harmonicevolution}. Two of the configurations ( \ref{5harmonicevolution}(c) and \ref{5harmonicevolution}(d)), are identical. They account for the case where only one of the two daughter loops at height $h=2$ self-intersects. Since labelling the loops is not of importance in our calculations, we can account for these two configurations as one configuration of multiplicity two. The probability of configuration (a) of plot \ref{5harmonicevolution} to occur is $P(5,5)$. The probability of configuration (b) is the probability of the 5 order loop to self-intersect, and none of the daughter loops to do so, which corresponds to $P(5,3)\left[1-P(3,1)\right]^2$. Using the same reasoning, the probability of configurations (c) and (d) is $P(5,3)$ $P(3,1)$ $\left[1-P(3,1)\right]$, and of (e) is $P(5,3)P(3,1)^2$.

We find that the average number of stable loops is
\begin{equation}
\begin{aligned}
n_s^{(5)}=&P(5,5)+2P(5,3)\left[1-P(3,1)\right]^2+6P(5,3)P(3,1)\left[1-P(3,1)\right]+\\
&4P(5,3)P(3,1)^2=2.76.
\label{stableloops5}
\end{aligned}
\end{equation}

The total number of cusps per period produced on average from the configuration (a) of plot \ref{5harmonicevolution} coincides with $c_5$ from Table 1 of \cite{Pazouli:2020qmj}, similarly to the $N=3$ case. Its lifetime is also the same as the lifetime of an $N=5$ string loop that does not self-intersect, $\alpha t_i/\Gamma G\mu$. The total number of cusps on average produced from configuration (b) is 
\begin{equation}
c_5 \frac{\frac{\alpha t_i}{4}}{\frac{\alpha t_i}{2}}+2c_3\frac{\frac{\alpha t_i}{2\Gamma G\mu }}{\frac{\alpha t_i}{4}}=\frac{1}{2}c_5+\frac{4c_3}{\Gamma G\mu}\simeq \frac{4c_3}{\Gamma G\mu}.
\label{conf5bc}
\end{equation}
The configurations (c) and (d) each produce a total number of  cusps per period
\begin{equation}
c_5 \frac{\frac{\alpha t_i}{4}}{\frac{\alpha t_i}{2}}+c_3\frac{\frac{\alpha t_i}{8}}{\frac{\alpha t_i}{4}}+c_3\frac{\frac{\alpha t_i}{2\Gamma G\mu }}{\frac{\alpha t_i}{4}}+2c_1\frac{\frac{\alpha t_i}{4\Gamma G\mu }}{\frac{\alpha t_i}{8}}\simeq \frac{2c_3+4c_1}{\Gamma G\mu}.
\label{conf5cdc}
\end{equation}
Finally, the configuration (e) produces the following total number of cusps per period 
\begin{equation}
c_5 \frac{\frac{\alpha t_i}{4}}{\frac{\alpha t_i}{2}}+2c_3\frac{\frac{\alpha t_i}{8}}{\frac{\alpha t_i}{4}}+4c_1\frac{\frac{\alpha t_i}{4\Gamma G\mu }}{\frac{\alpha t_i}{8}}\simeq \frac{8c_1}{\Gamma G\mu}.
\label{conf5ec}
\end{equation} 
Given the above, we find that an $N=5$ loop, which is allowed to self-intersect, will produce on average
\begin{equation}
\begin{aligned}
&P(5,5)c_5+P(5,3)\left[1-P(3,1)\right]^2 2c_3+2P(5,3)P(3,1)\\
&\left[1-P(3,1)\right](c_3+2c_1)+4P(5,3)P(3,1)^2c_1=9.7
\label{conf5c}
\end{aligned}
\end{equation}
cusps per period of the initial loop.

\subsection{Implementation with Mathematica}\label{ImplMath}
The flow chart of the method we used to calculate higher order harmonic cases is the following: 

\begin{enumerate}
\item Set the value of ``treeheight" (which is equal to the longest tree height minus one) to the value that we are interested in calculating. Note that the harmonic order of the parent loop $K$ is related to the value of ``treeheight" according to
\begin{equation}
K=2\, \text{``treeheight"}+3
\label{}
\end{equation}
Then, we initialize the value of a list for the tree of height 0 and the tree of height 1, which will be needed to calculate trees of higher height. This list is defined to include the following elements; the tree height, the number of leaves, the number of equivalent trees with the given characteristics, the probability of each tree configuration, and a list called ``newconttreedata" which contains all the information for the structure of the tree, i.e. its internal nodes and leaves at every level of the tree, and calculates their cusp contribution according to equation \eqref{ctildekl}. We also need a method to produce the trees of the next level, given that we have calculated all possible trees of a given level. For this, we use a list ``conttree", which contains all the information needed for this purpose. We initialize the value of ``conttree" for the smallest value (i.e. for transitioning from the level 0 tree to 1). The list includes the probability of the parent loop spliting, which is $P(K, K - 2)$, and its cusp contribution. We also initialize the list ``AllContTrees", which contains all trees used to produce next level trees, and the list ``AllTrees", which contains all trees.   

\item For the values between 1 and ``treeheight" repeat the following 
\begin{enumerate}
\item Initialize the list that saves the new trees types, produced in this iteration. Also, set the number of tree categories, given by the length of ``AllContTrees".
\item For all the tree types previously found (i.e. up to the previously calculated trees) repeat the following
\begin{enumerate}
\item Set the variables (of the given tree category) for the height ``h", for the number of total leaves (``leaves"), for the number of leaves at the bottom layer (``bleaves"), for the number of trees of this type (``n") and the probability of this tree type (``P"). Also keep in a list (called ``conttreedata") the number of leaves and internal nodes at every level of this type of tree.
\item For all the possible values of leaves on the bottom layer (i.e. values between 1 and ``bleaves") repeat the following
\begin{enumerate}
\item For the value of the bottom leaves ``i" of the previous tree, calculate the values of ``h", ``leaves", ``bleaves", ``n" and ``P" and the information for leaves and internal nodes for the new tree type (the list ``newconttreedata"). The height of the new tree will be $(h+1)$, the number of bottom leaves will be $2i$, the number of total leaves will be $\text{leaves}+1$, the number of trees of this type will be
\begin{equation}
n{\text{bleaves}\choose i}.
\label{}
\end{equation}
Note that in the above we have used the binomial coefficient
\begin{equation}
{n\choose k}=\frac{n!}{k!(n-k)!},
\label{}
\end{equation}
defined for positive integer values of $n$ and $k$. The probability of the new tree is
\begin{equation}
\begin{aligned}
&P \times P(K-2h,K-2h)^{\text{bleaves}-i}\times\\
& P(K-2h,K-2h-2)^i \times\\ 
&P(K-2h-2,K-2h-2)^{2i},
\label{}
\end{aligned}
\end{equation}
where $P$ is the probability of the tree type of ``treeheight"$=h$. We also create the list used to create the $h+2$ trees.
\item Add the above calculated tree to the list of trees.
\end{enumerate}
\end{enumerate}
\end{enumerate}
\item Use the formulas \eqref{averstloseries} - \eqref{slm} to calculate the average number of stable loops and the formulas \eqref{gkl} - \eqref{cfinalk} to calculate the average number of cusps.
\item Print the results analytically and then the numerical values by using the specified values for the cusps per period from Table I in \cite{Pazouli:2020qmj} and the probabilities $P(M,M-2)$ from Table \ref{tableselfinter}.
\end{enumerate}

Note that we also use the Monte Carlo method to test numerically the above results. We find the same values with both methods.

\bibliography{References}{}
\bibliographystyle{unsrt}

\end{document}